\pdfoutput=1

\documentclass[11pt]{article}

\usepackage[margin=1in]{geometry}
\usepackage[round,authoryear]{natbib}
\usepackage{times}
\usepackage{latexsym}
\usepackage[T1]{fontenc}
\usepackage[utf8]{inputenc}
\usepackage{microtype}
\usepackage{inconsolata}
\usepackage{graphicx}
\usepackage{amssymb}
\usepackage{amsmath}
\usepackage{booktabs}
\usepackage{subcaption}
\usepackage{tabularx}
\usepackage{multirow}
\usepackage{makecell}
\usepackage{xcolor}
\usepackage{rotating}
\usepackage{float}
\usepackage{stfloats}
\usepackage[bottom]{footmisc}
\usepackage[hidelinks]{hyperref}

\expandafter\def\expandafter\normalsize\expandafter{%
    \normalsize%
    \setlength\abovedisplayskip{4pt}%
    \setlength\belowdisplayskip{4pt}%
    \setlength\abovedisplayshortskip{0pt}%
    \setlength\belowdisplayshortskip{2pt}%
}

\newcommand{\mask}{\texttt{[MASK]}}
\newcommand{\promptreps}{PromptReps}

\newcommand{\method}{DiffRetriever}

\title{DiffRetriever: Parallel Representative Tokens for Retrieval with Diffusion Language Models}

\author{%
	\begin{tabular}{ccccc}
		Shuai Wang\textsuperscript{1} &
		Yu Yin\textsuperscript{1} &
		Shengyao Zhuang\textsuperscript{1} &
		Bevan Koopman\textsuperscript{1,2} &
		Guido Zuccon\textsuperscript{1}
	\end{tabular}\\
	\textsuperscript{1}The University of Queensland, Brisbane, QLD, Australia
	\quad
	\textsuperscript{2}CSIRO\\
	\texttt{\{shuai.wang2, y.yin1, s.zhuang, g.zuccon\}@uq.edu.au}
	\quad
	\texttt{bevan.koopman@csiro.au}
}
\date{}

\begin{document}
\maketitle

\begin{abstract}
	This paper shows how diffusion language models (DLMs) can be used as effective and efficient retrievers. Existing DLM-based retrievers (e.g., DiffEmbed) follow BERT-style encoding, representing each query or passage as a single mean-pooled vector. This ignores how DLMs are trained to generate responses through masked-position prediction under bidirectional attention, a capability that can provide stronger retrieval signals.
	We propose \method{}, which uses the DLM's native masked-position prediction directly for retrieval. For each query or passage, \method{} appends one or more masked positions, using the outputs as retrieval representations in a single forward pass. With one masked position, single-representation \method{} already improves over DiffEmbed on the same backbones. 
	\method{} also naturally extends to multi-representation retrieval: DLMs process multiple masked positions jointly, enabling ColBERT-style fine-grained matching with little additional encoding latency. In autoregressive LLM retrievers, the same multi-representation strategy requires sequential decoding and therefore incurs much higher latency.
	\method{} obtains the strongest aggregate effectiveness within our matched comparison, outperforming DiffEmbed, \promptreps{}, and RepLLaMA. Masked-position counts selected on training data transfer well across datasets, while per-query variation suggests headroom for adaptive allocation. Code is available at \url{https://github.com/ielab/diffretriever}.
\end{abstract}

\begin{center}
	\vspace{0.6em}
	\includegraphics[width=0.65\textwidth]{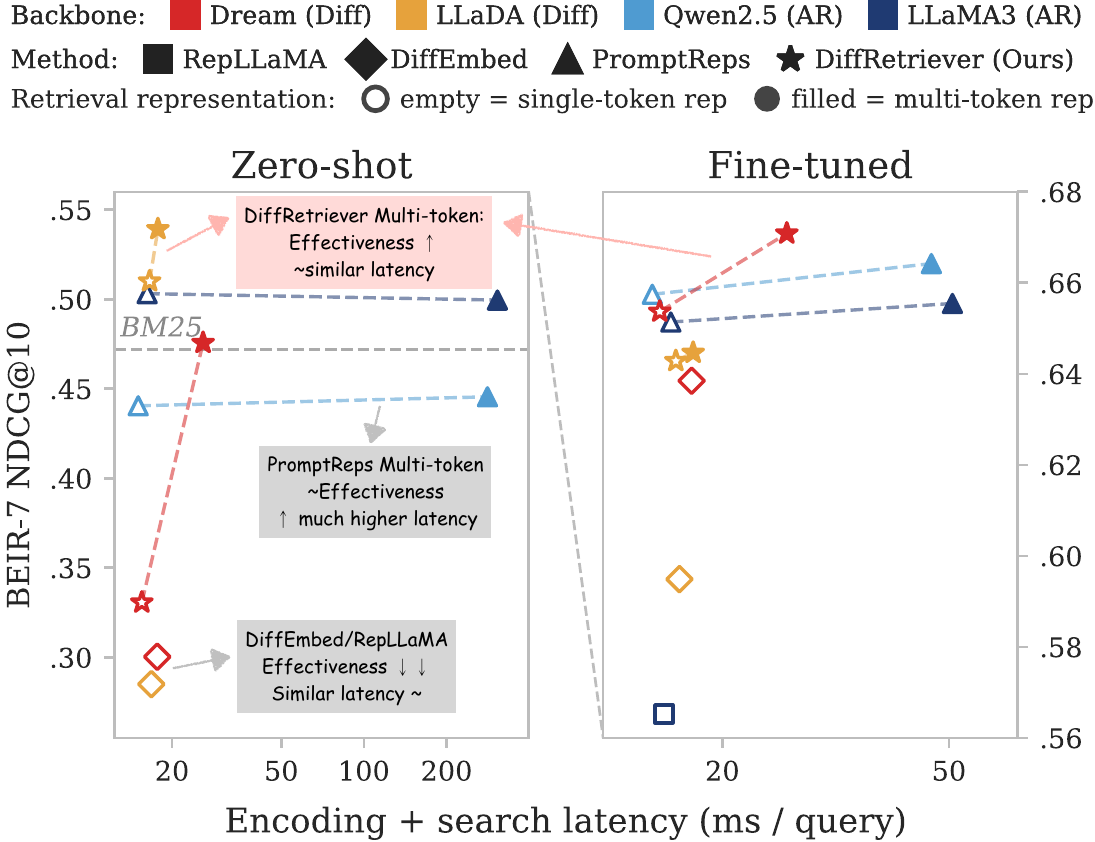}
	\captionof{figure}{BEIR-7 NDCG@10 vs.\ query encoding plus search latency
		(ms/query, $100$K-document sample). Appendix~\ref{app:latency_setup}
		reports full latency scaling.}
	\label{fig:teaser}
\end{center}

\section{Introduction}
\label{sec:introduction}
Diffusion language models (DLMs) generate responses through a masked-position prediction interface: they allocate masked output positions in advance and predict them jointly under bidirectional attention~\citep{nie2025large,ye2025dream}. Unlike autoregressive language models, which generate one token at a time, strictly left to right, DLMs can process several masked positions jointly.

Existing DLM-based retrievers do not take advantage of this capability. Methods such as DiffEmbed~\citep{zhang2025diffusion} follow BERT-style encoding of query or passage into a single mean-pooled vector for similarity scoring. As a result, retrieval does not benefit from the masked-position prediction DLMs were trained to perform, yielding suboptimal retrieval effectiveness.

We propose a DLM retriever that directly exploits masked-position prediction for retrieval. For each query or passage, \method{} prompts the model to represent the input, appends one or more masked positions after the prompt, and uses the resulting representations at those positions for retrieval. This makes \method{} a natural multi-representation retriever: increasing the number of allocated masked positions increases the number of  retrieval representations without requiring sequential generation.

Multi-representation retrievers were first explored with BERT-based models such as ColBERT, which proved more effective than single-representation retrieval by allowing fine-grained matching between query and passage tokens~\citep{khattab2020colbert,santhanam-etal-2022-colbertv2}. The same idea has been tested for autoregressive LLM retrievers, but did not yield consistent effectiveness gains while incurring extra cost of generating retrieval tokens sequentially~\citep{zhuang2024promptreps}. This raises a question: are multiple representations ineffective for LLM retrievers, or is sequential autoregression poorly suited to producing them? Our results point to the latter: the interface used to produce retrieval representations matters.

In both zero-shot and fine-tuned evaluation (Figure~\ref{fig:teaser}), even with a single masked position, \method{} already improves over DiffEmbed on the same backbones, showing the value of using the masked-position prediction interface directly. Multi-representation \method{} adds further gains with little additional latency, while comparable autoregressive retrievers incur much higher latency and obtain no consistent gain from multiple representations. Across MS~MARCO dev, TREC DL 19/20, and BEIR-7, \method{} achieves the strongest aggregate effectiveness within our matched comparison, outperforming DiffEmbed, \promptreps{}, and RepLLaMA. 
We further analyze where the gains come from by separating the effects of multiple masked-position representations and ColBERT-style MaxSim scoring. Multiple representations help even when simply mean-pooled, and MaxSim scoring adds further gains. We also study the impact of the number of masked-position tokens for representation: optimal values from training data transfer well across datasets, but per-query analysis shows significant gains by selecting an optimal number of masked tokens. Simple features such as query length and entropy correlate with this variation, suggesting adaptive masked-position prediction as a promising direction.

\section{Related Work}
\label{sec:related_work}

\paragraph{Diffusion language models in retrieval.}
Diffusion language models such as Dream~\citep{ye2025dream} and LLaDA~\citep{nie2025large} have motivated several retrieval methods built on DLMs, but existing methods do not use their masked-position prediction interface for retrieval. DiffEmbed~\citep{zhang2025diffusion} and pplx-embed~\citep{eslami2026diffusion} treat the DLM as a BERT-style encoder, producing a single mean-pooled representation from bidirectional hidden states. DiffuRank~\citep{liu2026diffurank} instead uses diffusion likelihood to rerank candidates from a first-stage retriever. In contrast, \method{} uses the masked-position prediction interface directly, producing one or more retrieval representations in a single bidirectional forward pass.

\paragraph{LLM-based and multi-representation retrieval.}
Autoregressive LLM-based embedding models and retrievers such as RepLLaMA~\citep{ma2024fine}, E5-Mistral~\citep{wang2024improving}, and GTE-Qwen~\citep{li2023towards} use contrastive fine-tuning to produce single-vector representations for first-stage retrieval. Recent general-purpose embedding models further improve retrieval through larger backbones and broader retrieval-oriented training data~\citep{muennighoff2024generative,lee2024nv,qwen3embedding}. \promptreps{}~\citep{zhuang2024promptreps} instead prompts autoregressive LLMs to generate a short textual representation and uses the corresponding hidden state and output logits as dense and sparse retrieval representations. \promptreps{} explored multi-representation variants, but found no reliable effectiveness gain over single-representation settings despite incurring sequential decoding cost.
Multi-representation retrieval has been most clearly established in BERT-based late-interaction models such as ColBERT~\citep{khattab2020colbert,santhanam-etal-2022-colbertv2}, where multiple representations per query and passage improve effectiveness through fine-grained matching. \method{} revisits this idea for LLM-based retrieval through the DLM masked-position interface: multiple representations are allocated in advance and produced jointly, enabling late interaction without autoregressive sequential decoding.

\section{Method}
\label{sec:method}
\begin{figure}[t]
	\centering
	\includegraphics[width=\textwidth]{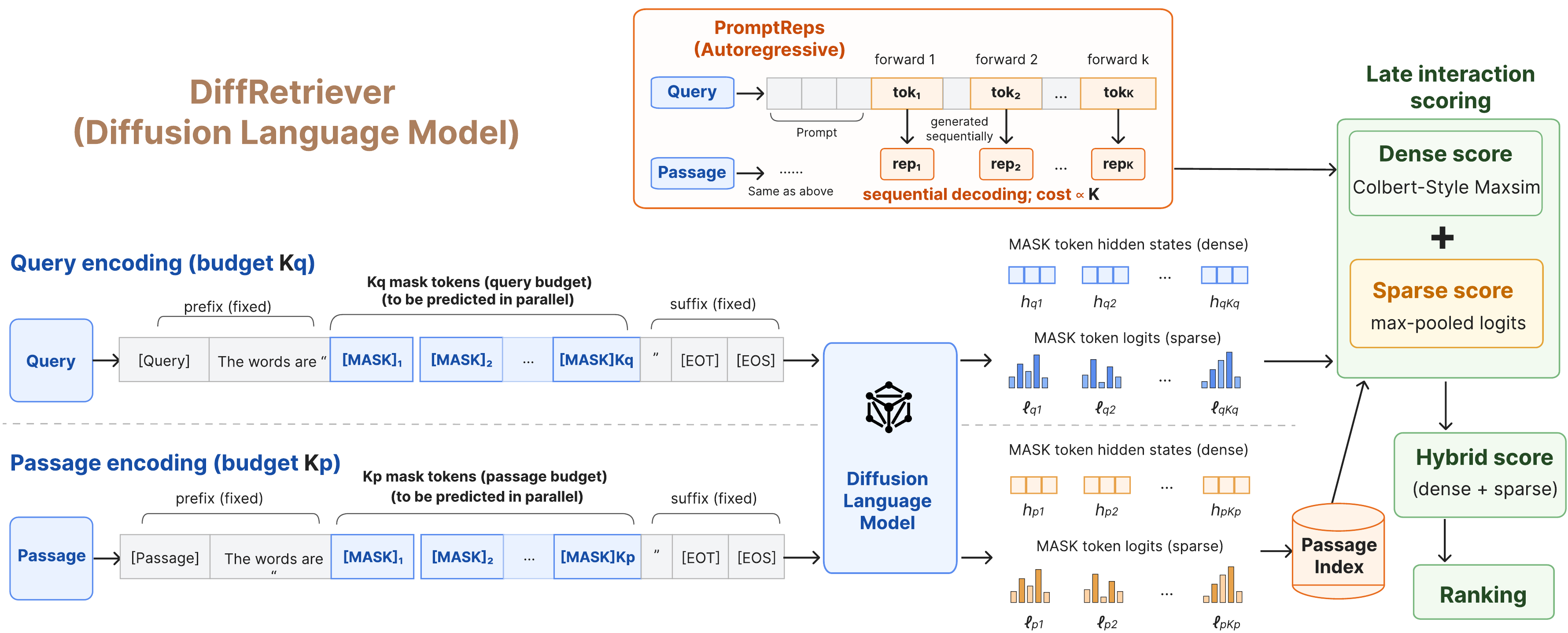}
	\caption{Overview of \method{}. A query and a passage are each
		formatted with a retrieval prompt that ends in a sequence of
		$K_q$ (query) or $K_p$ (passage) \mask{} positions, capped by
		fixed suffix tokens. A diffusion language model reads all masked
		positions in a single bidirectional forward pass, yielding $K$
		hidden states for dense retrieval and $K$ next-token logit
		vectors for sparse retrieval on each side. Scoring uses
		ColBERT-style MaxSim on the dense vectors and max-pooled logits
		on the sparse vectors, followed by hybrid score interpolation for
		ranking. Top-left inset: \promptreps{} on an autoregressive
		backbone generates retrieval representations sequentially up to a
		maximum budget $N$, so wall-clock encoding cost grows with the
		number generated; \method{} pre-allocates $K$ masked positions
		and reads them in parallel from one pass.}
	\label{fig:architecture}
\end{figure}

\method{} uses masked positions in DLMs as retrieval representations. Given a query or passage, it applies a retrieval prompt and appends $K$ masked positions, where $K{=}1$ gives the single-representation setting and $K{>}1$ gives the multi-representation setting. The resulting outputs support dense, sparse, and hybrid retrieval, and the same pipeline is used for supervised contrastive fine-tuning. Figure~\ref{fig:architecture} gives an end-to-end view.

Figure~\ref{fig:architecture} shows the contrast with autoregressive
\promptreps{} explicitly: \promptreps{} produces retrieval
representations one token at a time up to a cap $N$, while
\method{} pre-allocates the representation budget and obtains all
masked-position representations from one bidirectional pass. The same
masked-position outputs provide both hidden states for dense retrieval
and logits for sparse retrieval, so dense, sparse, and hybrid scoring
share a single encoding pipeline.

\subsection{Retrieval Prompt}
\label{sec:method_prompt}

Table~\ref{tab:prompt} shows the retrieval prompts for queries and
passages. The prompt frames the task for the model in two parts: a
system message that sets up an assistant role, and a user message that
asks the model to produce a textual representation of the input for
retrieval. The number of words requested controls the number of
retrieval representations: \textit{``one word''} for the
single-representation case ($K{=}1$), and \textit{``a few words''} for
the multi-representation case ($K{>}1$).\footnote{The original
\promptreps{} multi-representation prompt used \textit{``three
words''}; we find \textit{``a few words''} performs better. See
Appendix~\ref{app:prompts}.}

\begin{table}[t]
	\centering
	\small
	\caption{Retrieval prompts for queries and passages. Square brackets indicate alternatives for single-representation ($K{=}1$) and multi-representation ($K{>}1$) cases.}
	\begin{tabularx}{\textwidth}{p{0.12\textwidth}XX}
		\toprule
		\textbf{Role} & \textbf{Query} & \textbf{Passage} \\
		\midrule
		System &
		You are an AI assistant that can understand human language. &
		You are an AI assistant that can understand human language. \\
		\midrule
		User &
		Query: ``$x$''. Use [\textbf{one word} $\mid$ \textbf{a few words}] to represent the query in a retrieval task. Make sure your [\textbf{word is} $\mid$ \textbf{words are}] in lowercase. &
		Passage: ``$x$''. Use [\textbf{one word} $\mid$ \textbf{a few words}] to represent the passage in a retrieval task. Make sure your [\textbf{word is} $\mid$ \textbf{words are}] in lowercase. \\
		\midrule
		Assistant &
		The [\textbf{word is} $\mid$ \textbf{words are}] `` &
		The [\textbf{word is} $\mid$ \textbf{words are}] `` \\
		\bottomrule
	\end{tabularx}
	\label{tab:prompt}
\end{table}

\subsection{Parallel Masked-Position Decoding}
\label{sec:method_schedule}

We use the DLM masked-position prediction interface directly for retrieval. Given a retrieval prompt, we append $K$ \mask{} positions followed by the closing tokens of the assistant response, and pass the full sequence through the model in a single forward pass:
\begin{equation}
	\begin{aligned}
		& \texttt{[chat prompt]}\;\texttt{The words are \bf{"}}                  \\
		& \quad \mask{}_1 \cdots \mask{}_K\;\texttt{\bf{"}}\;
		\texttt{<turn-end>}\;\texttt{<eos>}.
	\end{aligned}
	\label{eq:diffusion_input}
\end{equation}
The closing quote and termination tokens make the masked span a complete, length-specified assistant response. This preserves the chat format seen during DLM fine-tuning while specifying how many masked positions the diffusion backbone should predict. Here, \texttt{<turn-end>} is the chat template's end-of-turn token (e.g., \texttt{<|im\_end|>} for Dream and \texttt{<|eot\_id|>} for LLaDA), and \texttt{<eos>} is the end-of-sequence token.

For each \mask{} position, we use the hidden state and logits as retrieval signals. This yields one retrieval representation per masked position and therefore $K$ representations for the input text. Because bidirectional attention lets each \mask{} position attend to the prompt, the suffix tokens, and the other masked positions, the representations are produced jointly rather than sequentially. Query and passage inputs may use different numbers of masked positions; we denote these counts by $K_q$ and $K_p$.

\paragraph{Single-step decoding.}
DLMs are trained as iterative denoisers: at inference, they fill masked positions over $S$ steps, revealing a subset per step and re-encoding the remaining masked positions under bidirectional attention. We report all main experiments at $S{=}1$, i.e., a single forward pass. Appendix~\ref{app:multistep} shows that iterative denoising with $S{>}1$ is significantly worse than $S{=}1$ at zero-shot across all DLM backbones we test, and gives mixed results after fine-tuning. This suggests that the retrieval gain comes from bidirectional processing of the appended masked positions, not from the iterative denoising procedure itself.

\paragraph{Comparison to \promptreps{}}~\citep{zhuang2024promptreps}.
The same retrieval prompt can also be used with an autoregressive LLM like \promptreps{}. Unlike DLM decoding, where $K$ masked positions are pre-allocated before encoding, autoregressive decoding generates output tokens one-at-a-time under a causal mask until either a closing quote is produced or a generation cap $N$ is reached. Thus the actual number of generated representations varies by input and is at most $N$. The retrieval representations are otherwise analogous: each generated token provides a hidden state and logits used for retrieval. The difference is whether these representations are produced in parallel or sequentially.\footnote{In FLOPs, diffusion has no asymptotic advantage over autoregression with KV caching: both scale linearly in the number of representations. The advantage is in wall-clock latency, since autoregressive forward passes must run sequentially while diffusion finishes in one parallel pass.}

\subsection{Scoring}
\label{sec:method_scoring}

For an input text $x$ with $K_x$ masked positions, each representation $k \in \{1,\ldots,K_x\}$ consists of a hidden state $\mathbf{h}_k \in \mathbb{R}^{H}$ and a logit vector $\boldsymbol{\ell}_k \in \mathbb{R}^{|V|}$. We use the hidden states to compute a dense score, the logits to compute a sparse score, and combine the two into a hybrid score by linear interpolation.

\paragraph{Dense.}
We score a query against a passage by matching each of the $K_q$ query representations to its closest of the $K_p$ passage representations and averaging the resulting similarities~\citep{khattab2020colbert}:
\begin{equation}
	s_{\text{dense}}(q, p) = \frac{1}{K_q}\sum_{i=1}^{K_q}
	\max_{j=1}^{K_p} \mathbf{h}_i^{q\,\top} \mathbf{h}_j^{p},
	\label{eq:maxsim}
\end{equation}
where $\mathbf{h}_i^q$ and $\mathbf{h}_j^p$ are the $i$-th query and $j$-th passage hidden states. This naturally handles unequal $K_q$ and $K_p$ and reduces to a standard inner product when $K_q{=}K_p{=}1$.

\paragraph{Sparse.}
For each logit vector $\boldsymbol{\ell}_k$, we apply $\log(1{+}\mathrm{ReLU}(\cdot))$ and aggregate across the $K_x$ representations by element-wise max-pooling over the vocabulary dimension:
\begin{equation}
	\begin{aligned}
		\mathbf{s}[v]
		&= \max_{k=1}^{K_x}
		\log\bigl(1 + \mathrm{ReLU}(\boldsymbol{\ell}_k[v])\bigr), \\
		&\hspace{1.5em} v \in \{1,\ldots,|V|\}.
	\end{aligned}
	\label{eq:sparse}
\end{equation}
The sparse score is the inner product $s_{\text{sparse}}(q, p) = \mathbf{s}_q^{\top}\mathbf{s}_p$. We apply the content-word filter from \promptreps{} unchanged, which restricts the sparse vocabulary to lowercase content tokens, removing stopwords, punctuation, and subword fragments before scoring.

\paragraph{Hybrid.}
We combine dense and sparse scores by equal-weight linear interpolation after min-max normalization within each retriever's top-1000 list, following prior dense--sparse interpolation work~\citep{wang2021bert,li2022interpolate}:
\begin{equation}
	s_{\text{hybrid}}(q, p) = \tfrac{1}{2}\,
	\tilde{s}_{\text{dense}}(q, p) + \tfrac{1}{2}\,
	\tilde{s}_{\text{sparse}}(q, p),
	\label{eq:hybrid}
\end{equation}
where $\tilde{s}$ denotes the normalized score.

\subsection{Supervised Fine-Tuning}
\label{sec:method_finetune}

We use standard contrastive fine-tuning to train each backbone for retrieval. Positive query-passage pairs are encouraged to score higher than negatives under both the dense and sparse scores. The same objective is used for all backbones. For each query $q$, let $p^{+}$ be a positive passage and $\mathcal{P}$ a candidate pool containing $p^{+}$, sampled hard negatives, and in-batch passages from other queries. The dense loss is InfoNCE with temperature $\tau$:
\begin{equation}
	\mathcal{L}_{\text{dense}} = -\log
	\frac{\exp\bigl(s_{\text{dense}}(q, p^{+}) / \tau\bigr)}
	{\sum_{p \in \mathcal{P}}
		\exp\bigl(s_{\text{dense}}(q, p) / \tau\bigr)}.
\end{equation}
The sparse loss $\mathcal{L}_{\text{sparse}}$ is the analogous InfoNCE loss on $s_{\text{sparse}}$, without temperature scaling (equivalently, $\tau{=}1$). The final objective is $\mathcal{L} = \mathcal{L}_{\text{dense}} + \mathcal{L}_{\text{sparse}}$. We use the same $(K_q, K_p)$ at training and inference time, so each backbone is trained and evaluated with the same numbers of query and passage masked positions. Full training details are given in \S\ref{sec:exp_finetune}.

\section{Experimental Setup}
\label{sec:experimental_setup}

\subsection{Models}
\label{sec:exp_models}

We compare two autoregressive and two diffusion LLM backbones
at similar parameter scale ($7$ to $8$B). The autoregressive
models are LLaMA3-8B-Instruct~\citep{grattafiori2024llama} and
Qwen2.5-7B-Instruct~\citep{qwen2.5}; the diffusion models are
Dream-v0-Instruct-7B~\citep{ye2025dream} and
LLaDA-8B-Instruct~\citep{nie2025large}. We refer to these as
LLaMA3, Qwen2.5, Dream, and LLaDA.

We pair each diffusion backbone with a similar autoregressive model that helps isolate the impact of decoding strategy alone.
Dream was initialized from Qwen2.5 and then trained with
masked-position denoising, so Dream and Qwen2.5 share architecture
and initialization but differ after diffusion training; this gives
our closest comparison between autoregressive and diffusion backbones.
LLaDA is trained from scratch as a diffusion language model, without
an autoregressive checkpoint. We pair it with LLaMA3, the closest
autoregressive model in size and also LLaDA's direct competitor in
the original paper, as a complementary comparison without shared
initialization.

\subsection{Baselines}
\label{sec:exp_baselines}

We compare \method{} against four main baselines: (1) BM25 uses the
Pyserini~\citep{lin2021pyserini} default hyperparameters and index for
each dataset. (2) \promptreps{}~\citep{zhuang2024promptreps} uses
Qwen2.5 and LLaMA3 as directly comparable autoregressive LLM retrieval
baselines. (3) DiffEmbed~\citep{zhang2025diffusion} uses Dream and
LLaDA as encoder-style alternatives on the same DLM backbones,
mean-pooling over the input sequence without prompting or masked-position
prediction. (4) RepLLaMA~\citep{ma2024fine} uses LLaMA3 as a
contrastively fine-tuned single-vector retrieval baseline.
For \promptreps{}, DiffEmbed, and RepLLaMA, we re-train each baseline
with the same training data, optimizer, schedule, and adapter
configuration as \method{}, so fine-tuned comparisons are not confounded
by different training recipes.
Appendix~\ref{app:additional_comparison} reports a broader landscape
comparison without matched training recipes, including earlier BERT-based
retrievers (Contriever, TAS-B, ColBERT-v2) and strong recent LLM
embedding models (Qwen3-Embedding, NV-Embed, GritLM, and PPLX).
In this broader comparison, \method{} achieves the best in-domain
MS~MARCO dev result among all compared approaches. On BEIR-7 transfer,
\method{} outperforms earlier neural retrievers but trails the strongest
dedicated embedding models, which are trained on substantially more
diverse retrieval data than our MS~MARCO-only fine-tuning.

\subsection{Datasets and Metrics}
\label{sec:exp_data}

We evaluate on three benchmarks. (1) \textit{MS~MARCO} passage ranking~\citep{bajaj2016ms} is reported on the dev set with MRR@10. (2) \textit{TREC DL 2019} and \textit{TREC DL 2020}~\citep{craswell2019overview,craswell2020overview} are reported with NDCG@10. (3) \textit{BEIR-7} is a seven-dataset subset of BEIR~\citep{thakur2021beir} spanning diverse domains and tasks, which we use to measure out-of-domain transfer, comprising Natural Questions, HotpotQA, SciFact, TREC-COVID, FiQA, ArguAna, and Quora; we report NDCG@10. The seven datasets span open-domain QA, multi-hop QA, scientific fact verification, biomedical retrieval, financial QA, argument retrieval, and duplicate-question detection.

For latency comparisons in Figure~\ref{fig:teaser}, we report query encoding plus search time using the attention backend across backbones on a $100$K-document sample of the MS~MARCO corpus. Appendix~\ref{app:latency_setup} reports latency scaling across input length and index size. Appendix~\ref{app:index_storage} reports index-storage scaling, where multi-representation retrieval increases passage-index storage roughly with $K_p$; standard ColBERTv2 compression can reduce this by roughly $6\times$ with limited effectiveness loss.

\subsection{Selecting $K_q$ and $K_p$}
\label{sec:exp_budget}

\begin{figure}[t]
	\centering
	\includegraphics[width=0.4\textwidth]{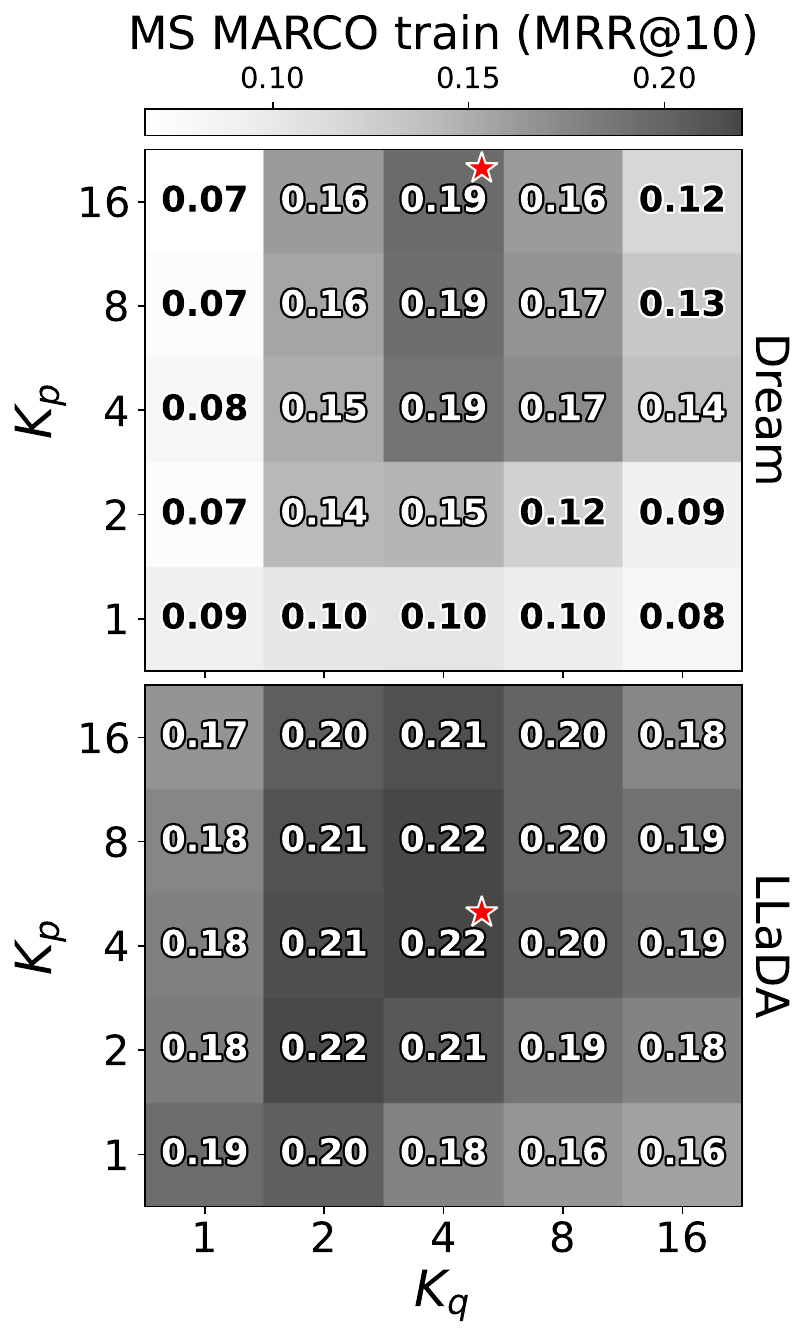}
	\caption{Zero-shot hybrid retrieval grid on MS~MARCO train,
		used for budget selection. Stars mark the argmax cell
		that defines the fixed $(K_q^{*}, K_p^{*})$ applied
		unchanged across all evaluations: $(4,16)$ for Dream and
		$(4,4)$ for LLaDA.}
	\label{fig:kqkp_heatmap_train}
\end{figure}

\method{} processes all appended \mask{} positions in
one forward pass, so $K_q$ and $K_p$ must be chosen before
encoding. For each diffusion backbone, we sweep
$(K_q,K_p) \in \{1,2,4,8,16\}^2$ on the MS~MARCO training set
and select the pair with the highest hybrid retrieval score,
allowing $K_q$ and $K_p$ to differ. This gives
$(K_q^{*},K_p^{*})=(4,16)$ for Dream and $(4,4)$ for LLaDA.
Each backbone then uses its selected $(K_q,K_p)$ unchanged across
all evaluations (MS~MARCO dev, TREC DL 2019/2020, and BEIR-7)
and as the train-time setting for supervised fine-tuning. Figure~\ref{fig:kqkp_heatmap_train}
shows the training grid used for this selection.

\subsection{Fine-Tuning Setup}
\label{sec:exp_finetune}

We fine-tune on the Tevatron MS~MARCO passage augmented
triples~\citep{gao2022tevatron,bajaj2016ms}, following
\promptreps{}. Each training item contains a query, one sampled
positive passage, and $15$ sampled hard negatives. Diffusion
backbones are fine-tuned at their selected $(K_q^{*},K_p^{*})$,
so training and inference use the same numbers of query and passage
masked positions. For \promptreps{}, we use a generation cap of
$N{=}4$ for both LLaMA3 and Qwen2.5 during fine-tuning.\footnote{Zero-shot uses
	$N{=}20$; we reduce the cap to $N{=}4$ during fine-tuning
	because larger caps exceed our memory budget at the global batch
	size used for matched fine-tuning.} Full fine-tuning hyperparameters are in
Appendix~\ref{app:finetune_details}.

\begin{table*}[t]
	\centering
	\small
	\setlength{\tabcolsep}{3pt}
	\caption{In-domain retrieval on MS~MARCO dev and TREC DL. D/S/H are dense, sparse, and hybrid scores; BM25 is sparse. Fwd is query-side forward steps. \textbf{Bold}: best within each block/column. $\dagger$: better than same-backbone single-rep; $\ddagger$: \method{} multi-rep better than corresponding \promptreps{} single-rep. Paired $t$-test, $p<0.05$.}
	\label{tab:main_indomain}
	\begin{tabular*}{\textwidth}{@{\extracolsep{\fill}}clllc ccc ccc ccc}
		\toprule
		& \multirow{2}{*}{Method} & \multirow{2}{*}{Backbone} & \multirow{2}{*}{Variant} & \multirow{2}{*}{Fwd} & \multicolumn{3}{c}{MS~MARCO dev} & \multicolumn{3}{c}{DL19} & \multicolumn{3}{c}{DL20} \\
		\cmidrule(lr){6-8} \cmidrule(lr){9-11} \cmidrule(lr){12-14}
		&        &          &         &     & D & S & H & D & S & H & D & S & H \\
		\midrule
		\multirow{11}{*}{\rotatebox[origin=c]{90}{\textbf{Zero-shot}}}
		& BM25                                & ---                            & sparse     & ---      & --- & .184 & --- & --- & \textbf{.506} & --- & --- & \textbf{.480} & --- \\
		\cmidrule(lr){2-14}
		& \multirow{2}{*}{DiffEmbed}          & Dream                          & single-rep & 1        & .034 & --- & --- & .184 & --- & --- & .110 & --- & --- \\
		&                                     & LLaDA                          & single-rep & 1        & .014 & --- & --- & .111 & --- & --- & .063 & --- & --- \\
		\cmidrule(lr){2-14}
		& \multirow{4}{*}{PromptReps}         & \multirow{2}{*}{Qwen2.5}       & single-rep & 1        & .100 & .190 & .200 & .320 & .407 & .439 & .245 & .428 & .422 \\
		&                                     &                                & multi-rep  & $\leq$20 & .142$^\dagger$ & .177 & .200 & .404 & .405 & .477 & .386$^\dagger$ & .410 & .478$^\dagger$ \\
		&                                     & \multirow{2}{*}{LLaMA3}        & single-rep & 1        & .178 & .211 & .242 & \textbf{.511} & .431 & .513 & .445 & .452 & \textbf{.536} \\
		&                                     &                                & multi-rep  & $\leq$20 & .151 & .196 & .220 & .319 & .454 & .473 & .336 & .412 & .471 \\
		\cmidrule(lr){2-14}
		& \multirow{4}{*}{\textbf{DiffRetriever}} & \multirow{2}{*}{Dream}         & single-rep & 1        & .086 & .073 & .112 & .325 & .313 & .369 & .255 & .208 & .312 \\
		&                                     &                                & multi-rep  & 1        & \textbf{.192}$^{\dagger\ddagger}$ & .165$^\dagger$ & .218$^{\dagger\ddagger}$ & .479$^{\dagger\ddagger}$ & .427$^\dagger$ & .496$^\dagger$ & .466$^{\dagger\ddagger}$ & .422$^\dagger$ & .512$^{\dagger\ddagger}$ \\
		&                                     & \multirow{2}{*}{LLaDA}         & single-rep & 1        & .158 & .200 & .221 & .391 & .451 & .481 & .367 & .463 & .483 \\
		&                                     &                                & multi-rep  & 1        & \textbf{.192}$^{\dagger\ddagger}$ & \textbf{.223}$^{\dagger\ddagger}$ & \textbf{.248}$^\dagger$ & .490$^\dagger$ & .479 & \textbf{.549}$^\dagger$ & \textbf{.472}$^\dagger$ & .479 & \textbf{.536}$^\dagger$ \\
		\midrule
		\multirow{11}{*}{\rotatebox[origin=c]{90}{\textbf{Fine-tuned}}}
		& RepLLaMA                            & LLaMA3                         & single-rep & 1        & .412 & --- & --- & .715 & --- & --- & .690 & --- & --- \\
		\cmidrule(lr){2-14}
		& \multirow{2}{*}{DiffEmbed}          & Dream                          & single-rep & 1        & .405 & --- & --- & .720 & --- & --- & .693 & --- & --- \\
		&                                     & LLaDA                          & single-rep & 1        & .398 & --- & --- & .695 & --- & --- & .676 & --- & --- \\
		\cmidrule(lr){2-14}
		& \multirow{4}{*}{PromptReps}         & \multirow{2}{*}{Qwen2.5}       & single-rep & 1        & .419 & .343 & .405 & .741 & .592 & .711 & .715 & .620 & .703 \\
		&                                     &                                & multi-rep  & $\leq$4  & .424 & .347 & .406 & .738 & .607 & .729 & .731 & \textbf{.632} & .705 \\
		&                                     & \multirow{2}{*}{LLaMA3}        & single-rep & 1        & .425 & .347 & .410 & .743 & .605 & .715 & \textbf{.751} & .631 & .707 \\
		&                                     &                                & multi-rep  & $\leq$4  & .430 & .348 & \textbf{.414} & .746 & .613 & .732 & .739 & .621 & \textbf{.709} \\
		\cmidrule(lr){2-14}
		& \multirow{4}{*}{\textbf{DiffRetriever}} & \multirow{2}{*}{Dream}         & single-rep & 1        & .424 & .341 & .405 & .741 & .614 & .724 & .721 & .627 & .690 \\
		&                                     &                                & multi-rep  & 1        & \textbf{.433}$^{\dagger\ddagger}$ & \textbf{.349}$^{\dagger\ddagger}$ & .411$^{\dagger\ddagger}$ & \textbf{.751} & .620 & \textbf{.739} & .729 & .617 & .697 \\
		&                                     & \multirow{2}{*}{LLaDA}         & single-rep & 1        & .424 & .347 & .405 & .715 & .621 & .704 & .715 & .624 & .701 \\
		&                                     &                                & multi-rep  & 1        & .427 & .348 & .408 & .718 & \textbf{.636} & .718 & .721 & .614 & .698 \\
		\bottomrule
	\end{tabular*}
\end{table*}

\section{Results}
\label{sec:results}

We report results in three settings: in-domain zero-shot
retrieval in Section~\ref{sec:results_zeroshot}, in-domain
fine-tuned retrieval in Section~\ref{sec:results_finetuned}, and
out-of-domain transfer to BEIR-7 in Section~\ref{sec:results_ood}.

\subsection{Zero-shot in-domain retrieval}
\label{sec:results_zeroshot}

Table~\ref{tab:main_indomain} (zero-shot rows) reports
zero-shot effectiveness. In the single-representation setting,
\promptreps{} achieves the highest effectiveness: LLaMA3 reaches
$.242$ hybrid MRR@10 on MS~MARCO dev, ahead of all \method{}
backbones. Dream-based \method{} is weakest in the
single-representation setting and falls below BM25 on MS~MARCO
dev. DiffEmbed, which uses the same DLM backbones as
mean-pooled encoders without masked-position prediction, performs
even worse, indicating that masked-position representations are
still preferable to encoder-style pooling even when the underlying
zero-shot DLM backbone is weak.

This pattern changes in the multi-representation setting. For
\promptreps{}, generating multiple representations yields no
consistent improvement: LLaMA3 hybrid drops on MS~MARCO dev and
DL19, Qwen2.5 shows no significant gain across all three
benchmarks, and 11 of the 18 \promptreps{} multi-representation
cells fall below the corresponding single-representation cells.
For \method{}, however, the multi-representation setting improves
effectiveness in every dataset and scoring mode; 16 of the 18
\method{} multi-representation cells are significantly better than
the corresponding single-representation cells. A scoring
decomposition in Section~\ref{sec:analysis_scoring} further shows
that the gain comes from both the representations and the scoring
function: mean-pooling the same multiple masked-position
representations already improves over the single-representation
setting, and MaxSim late-interaction scoring adds further gains.

To partially control for backbone differences, we compare
Dream-based \method{} with Qwen2.5-based \promptreps{}, since
Dream was initialized from Qwen2.5 before diffusion training.
Their ordering changes with the number of representations:
Qwen2.5 leads in the single-representation setting, whereas
multi-representation Dream leads in all cases except sparse
scoring on MS~MARCO dev. This suggests that the gains are tied
to the parallel masked-position interface rather than to a
diffusion-backbone advantage alone.

Latency makes the tradeoff sharper. In zero-shot evaluation,
\promptreps{} with multiple generated representations is capped at
$N=20$ and takes $275$--$300$~ms per query, whereas \method{}
takes $16$--$20$~ms (Figure~\ref{fig:teaser}). The roughly
$15\times$ latency gap, combined with the lack of consistent
multi-representation gains in \promptreps{}, suggests that the
bottleneck is sequential generation rather than multi-representation
retrieval itself.

\subsection{In-domain fine-tuning}
\label{sec:results_finetuned}
\begin{table*}[t]
	\centering
	\small
	\setlength{\tabcolsep}{3pt}
	\caption{Out-of-domain BEIR-7 NDCG@10. Zero-shot uses hybrid scores where available; fine-tuned uses dense scores. Fwd is query-side forward steps. \textbf{Bold}: best within each block/column. $\dagger$: better than same-backbone single-rep; $\ddagger$: \method{} multi-rep better than corresponding \promptreps{} single-rep. Paired $t$-test, $p<0.05$.}
	\label{tab:main_beir7}
	\begin{tabular*}{\textwidth}{@{\extracolsep{\fill}}clllccccccccc}
		\toprule
		&        &          &         &     & \multicolumn{8}{c}{BEIR-7 (NDCG@10)} \\
		\cmidrule(lr){6-13}
		& Method & Backbone & Variant & Fwd & NQ & HQA & SciFact & COVID & FiQA & ArguAna & Quora & Avg \\
		\midrule
		\multirow{11}{*}{\rotatebox[origin=c]{90}{\textbf{Zero-shot}}}
		& BM25                                & ---                            & sparse     & ---      & .243 & \textbf{.567} & \textbf{.664} & .530 & .236 & .275 & .789 & .472 \\
		\cmidrule(lr){2-13}
		& \multirow{2}{*}{DiffEmbed}          & Dream                          & single-rep & 1        & .118 & .149 & .383 & .316 & .143 & .308 & .685 & .300 \\
		&                                     & LLaDA                          & single-rep & 1        & .066 & .167 & .385 & .407 & .117 & .326 & .528 & .285 \\
		\cmidrule(lr){2-13}
		& \multirow{4}{*}{PromptReps}         & \multirow{2}{*}{Qwen2.5}       & single-rep & 1        & .302 & .322 & .618 & .577 & .271 & .209 & .784 & .441 \\
		&                                     &                                & multi-rep  & $\leq$20 & .316$^\dagger$ & .470$^\dagger$ & .609 & .538 & .239 & .240$^\dagger$ & .707 & .446$^\dagger$ \\
		&                                     & \multirow{2}{*}{LLaMA3}        & single-rep & 1        & .410 & .454 & .643 & .645 & .318 & .248 & .804 & .503 \\
		&                                     &                                & multi-rep  & $\leq$20 & .379 & .507$^\dagger$ & .639 & .613 & .270 & .316$^\dagger$ & .772 & .500 \\
		\cmidrule(lr){2-13}
		& \multirow{4}{*}{\textbf{DiffRetriever}} & \multirow{2}{*}{Dream}         & single-rep & 1        & .153 & .066 & .380 & .497 & .220 & .273 & .725 & .331 \\
		&                                     &                                & multi-rep  & 1        & .358$^{\dagger\ddagger}$ & .449$^{\dagger\ddagger}$ & .636$^\dagger$ & .590$^\dagger$ & .278$^\dagger$ & .323$^{\dagger\ddagger}$ & .696 & .476$^{\dagger\ddagger}$ \\
		&                                     & \multirow{2}{*}{LLaDA}         & single-rep & 1        & .382 & .299 & .618 & \textbf{.752} & \textbf{.325} & \textbf{.370} & \textbf{.825} & .510 \\
		&                                     &                                & multi-rep  & 1        & \textbf{.415}$^\dagger$ & .494$^{\dagger\ddagger}$ & .660$^\dagger$ & .717$^\ddagger$ & .317 & .357$^\ddagger$ & .814$^\ddagger$ & \textbf{.539}$^{\dagger\ddagger}$ \\
		\midrule
		\multirow{11}{*}{\rotatebox[origin=c]{90}{\textbf{Fine-tuned}}}
		& RepLLaMA                            & LLaMA3                         & single-rep & 1        & .622 & .503 & .653 & .564 & .383 & .411 & .821 & .565 \\
		\cmidrule(lr){2-13}
		& \multirow{2}{*}{DiffEmbed}          & Dream                          & single-rep & 1        & .625 & .668 & .735 & .758 & .478 & .375 & .831 & .638 \\
		&                                     & LLaDA                          & single-rep & 1        & .587 & .601 & .714 & .578 & .455 & .387 & .842 & .595 \\
		\cmidrule(lr){2-13}
		& \multirow{4}{*}{PromptReps}         & \multirow{2}{*}{Qwen2.5}       & single-rep & 1        & .619 & .661 & .756 & .843 & .444 & .407 & .872 & .657 \\
		&                                     &                                & multi-rep  & $\leq$4  & .623 & .686$^\dagger$ & \textbf{.773}$^\dagger$ & \textbf{.848} & .440 & .407 & .872 & .664$^\dagger$ \\
		&                                     & \multirow{2}{*}{LLaMA3}        & single-rep & 1        & .619 & .685 & .750 & .815 & .417 & .414 & .860 & .651 \\
		&                                     &                                & multi-rep  & $\leq$4  & .631$^\dagger$ & \textbf{.696}$^\dagger$ & .735 & .809 & .437$^\dagger$ & \textbf{.416} & .865$^\dagger$ & .655$^\dagger$ \\
		\cmidrule(lr){2-13}
		& \multirow{4}{*}{\textbf{DiffRetriever}} & \multirow{2}{*}{Dream}         & single-rep & 1        & .619 & .650 & .739 & .841 & .463 & .406 & .859 & .654 \\
		&                                     &                                & multi-rep  & 1        & \textbf{.644}$^{\dagger\ddagger}$ & .683$^{\dagger\ddagger}$ & .752 & .847 & \textbf{.479}$^{\dagger\ddagger}$ & .403 & \textbf{.887}$^{\dagger\ddagger}$ & \textbf{.671}$^{\dagger\ddagger}$ \\
		&                                     & \multirow{2}{*}{LLaDA}         & single-rep & 1        & .620 & .640 & .733 & .840 & .453 & .414 & .799 & .643 \\
		&                                     &                                & multi-rep  & 1        & .622 & .647$^\dagger$ & .744 & .846 & .443$^\ddagger$ & .412 & .798 & .645 \\
		\bottomrule
	\end{tabular*}
\end{table*}

After contrastive fine-tuning, multi-representation \method{} with
Dream is the strongest system within our matched comparison
(Table~\ref{tab:main_indomain}, fine-tuned rows), although
margins are smaller than in the zero-shot setting. It achieves the
highest score in four of the nine columns, and exceeds multi-representation \promptreps{},
DiffEmbed on the same Dream backbone, and RepLLaMA on MS~MARCO
dense retrieval. The same-backbone DiffEmbed comparison shows
that, under the same training recipe, masked-position
representations outperform encoder-style mean pooling.

The strongest scoring mode also shifts under supervision. Hybrid
scoring is best for every in-domain zero-shot configuration,
whereas dense scoring is usually strongest after fine-tuning. This
suggests that contrastive supervision mainly strengthens the dense
semantic signal, making equal-weight interpolation with the sparse
score less helpful. We therefore report hybrid scores for
zero-shot transfer and dense scores for fine-tuned transfer in the
remainder of the paper.

Dream and LLaDA respond differently to the same training recipe
under both DiffEmbed and \method{}. In the zero-shot setting,
LLaDA tends to be stronger; after fine-tuning, Dream becomes
stronger. One possible explanation is the difference in
initialization: Dream is initialized from Qwen2.5 and may inherit
representations shaped by next-token prediction, while LLaDA is
trained from scratch as a diffusion model. These inherited
representations may be better suited to the discriminative
matching objective used in contrastive fine-tuning. The same
asymmetry appears more sharply out of domain
(\S\ref{sec:results_ood}).

\subsection{Out-of-domain transfer}
\label{sec:results_ood}

Table~\ref{tab:main_beir7} reports out-of-domain transfer on
BEIR-7. Following the scoring-mode shift observed in
\S\ref{sec:results_finetuned}, we report zero-shot results with
hybrid scoring and fine-tuned results with dense scoring. The
full dense/sparse/hybrid breakdown is in
Appendix~\ref{app:beir7_full}.

\paragraph{Zero-shot transfer.}
The zero-shot pattern from in-domain evaluation generalizes to
BEIR-7. In the single-representation setting, \method{} does not
improve over \promptreps{}: Dream trails Qwen2.5, and LLaDA is
roughly level with LLaMA3. With multiple representations,
however, both comparisons reverse: Dream-based \method{} ($.476$)
beats Qwen2.5-based \promptreps{}, while LLaDA-based \method{}
($.539$) beats LLaMA3-based \promptreps{} and is the strongest
zero-shot BEIR-7 system in our comparison. \method{} substantially
outperforms DiffEmbed on the same diffusion backbones, suggesting
that masked-position representations are more effective than
encoder-style mean pooling for these models.

\paragraph{Fine-tuned transfer.}
The in-domain fine-tuned trend largely carries over to BEIR-7.
Multi-representation \method{} with Dream achieves the highest
BEIR-7 average in our comparison ($.671$), ahead of
multi-representation \promptreps{} with Qwen2.5 ($.664$) and
DiffEmbed with Dream ($.638$). The improvement is statistically
significant over DiffEmbed with Dream and over the
single-representation LLaMA3 \promptreps{} baseline.

A broader pattern is that fine-tuned transfer depends strongly on
the backbone family. Qwen2.5-family systems, namely \method{} with
Dream and \promptreps{} with Qwen2.5, transfer better than
LLaMA-based and LLaDA-based systems. Thus, out-of-domain
fine-tuned effectiveness reflects both the retrieval mechanism and
the transfer behavior of the underlying backbone.

\paragraph{Overall.}
Across both zero-shot and fine-tuned transfer, multi-representation
\method{} gives the best BEIR-7 average in our comparison while
retaining single-forward-pass encoding. The gains are larger with
Dream than with LLaDA, showing that the benefit of the diffusion
decoding interface also depends on the underlying backbone.

\section{Analysis}
\label{sec:analysis}

The results above show that multi-representation \method{} improves
retrieval, but we still cannot attribute the gain to either 1) multi-token representation via a diffusion model; or 2) late-interaction MaxSim scoring (following ColBERT). To rule out the effect of late interaction, we can simply apply mean-pooling to each model.

We then turn to the impact of the number of masked tokens $(K_q,K_p)$. Recall that this was fixed based on MS~MARCO train results and reused across all datasets. We investigate the impact of setting this to be dataset-specific.

\subsection{Where do the gains come from?}
\label{sec:analysis_scoring}

\begin{figure}[t]
	\centering
	\includegraphics[width=0.8\textwidth]{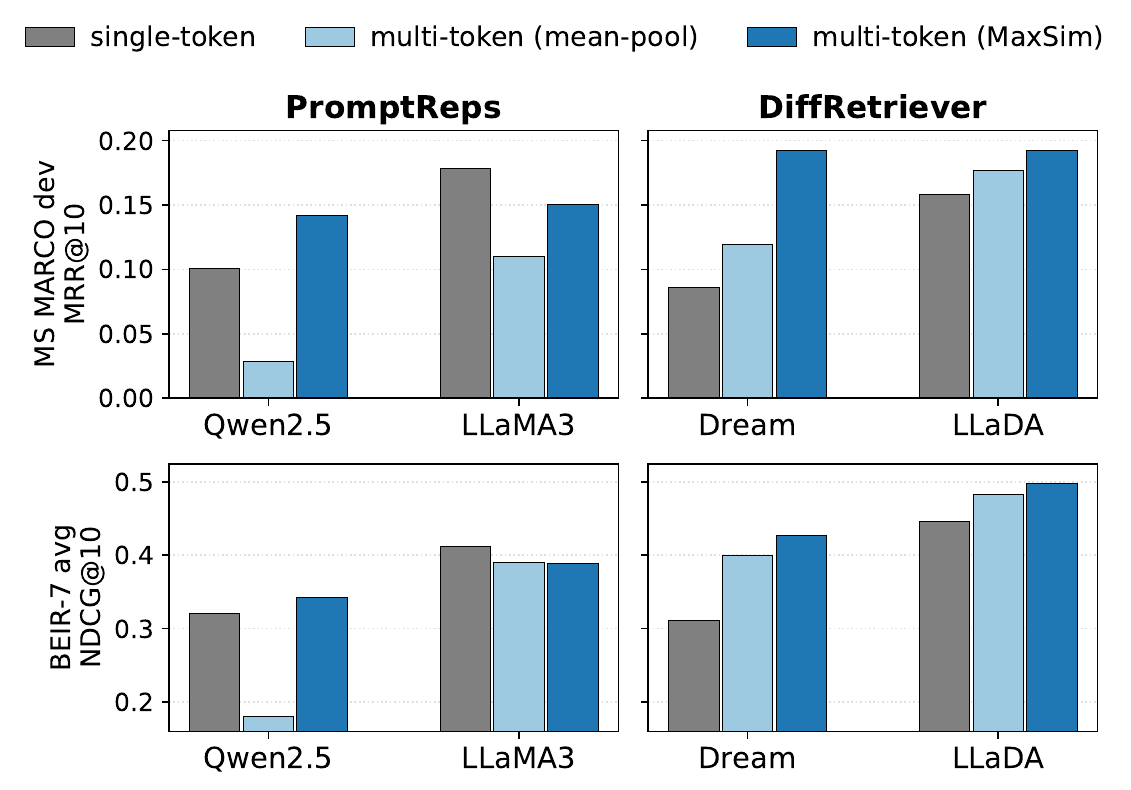}
	\caption{Dense scoring decomposition on MS~MARCO dev
		(MRR@10) and BEIR-7 average (NDCG@10).}
	\label{fig:scoring_decomposition}
\end{figure}

Multi-representation \method{} combines two changes: it uses
several masked-position representations and compares them with
ColBERT-style late interaction. Figure~\ref{fig:scoring_decomposition}
separates these effects by comparing three dense scoring variants:
a single representation, multiple representations with mean
pooling, and multiple representations with ColBERT MaxSim.

For \method{}, mean-pooling multiple masked-position representations
already improves over the single-representation setting on both
diffusion backbones. This shows that the added masked positions
provide useful retrieval information even before late interaction is
applied. MaxSim further adds gains, so late interaction is beneficial
but not the sole source of the improvement. In contrast, \promptreps{}
shows consistent declines from multiple generated representations under
mean pooling, matching the main finding that autoregressive
multi-representation retrieval pays higher decoding cost without
reliable effectiveness improvements.

\subsection{The impact of number of masked tokens}
\label{sec:analysis_landscape}

\begin{center}
	\centering
	\includegraphics[width=0.8\textwidth]{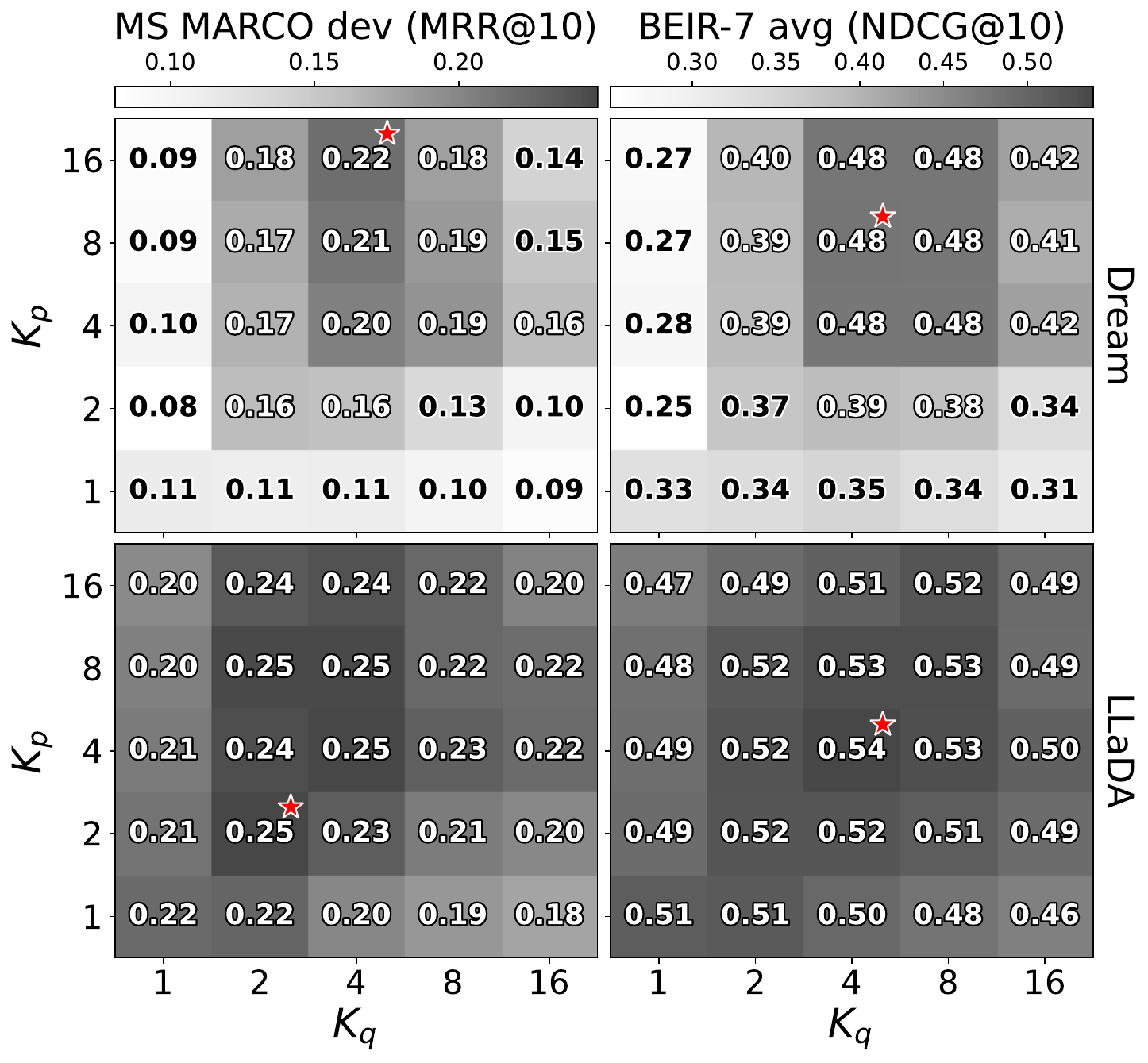}
	\captionof{figure}{Zero-shot hybrid retrieval across $(K_q,K_p)\in\{1,2,4,8,16\}^2$ on MS~MARCO dev and BEIR-7. Stars mark the best cell per panel; train-selected values, held fixed across evaluations, are $(4,16)$ for Dream and $(4,4)$ for LLaDA. Train grid: Figure~\ref{fig:kqkp_heatmap_train}.}
	\label{fig:kqkp_heatmap}
\end{center}

The main experiments select $(K_q,K_p)$ based on MS~MARCO training
and apply this to all datasets. This could be
fragile if the best masked-position counts varied sharply across
datasets. Figure~\ref{fig:kqkp_heatmap} shows that this is not
the case at the aggregate level. Dream selects an asymmetric,
passage-heavy setting, $(K_q^{*},K_p^{*})=(4,16)$, whereas LLaDA
selects the symmetric setting $(4,4)$. These train-selected cells
lie at or near the best test-set cell on both MS~MARCO dev and
the BEIR-7 average, and the per-backbone preference persists
across distributions.

However, the landscapes differ in sensitivity. On MS~MARCO dev,
LLaDA spans $\Delta=.07$ across grid cells, whereas Dream spans
$\Delta=.14$. Thus, fixed masked-position counts remain effective
in aggregate, but the cost of a poor choice is backbone-dependent and
larger for Dream.~\footnote{Appendix~\ref{app:per_dataset_heatmaps} reports results for the per-dataset optima.}
The aggregate robustness does not mean that a single fixed budget is
optimal for every query, which we analyze next.

\subsection{Per-query oracle headroom and predictability}
\label{sec:analysis_oracle}

The main experiments use one fixed representation budget
$(K_q^{*},K_p^{*})$ per diffusion backbone. This is the deployable
setting used throughout the paper, and Appendix~\ref{app:per_dataset_heatmaps}
shows that these fixed budgets transfer well in aggregate. We next ask
how much effectiveness could be gained if the budget were chosen
separately for each query, and whether query-side budget preferences
can be predicted from cheap pre-encoding features.

\begin{figure}[t]
	\centering
	\includegraphics[width=\textwidth]{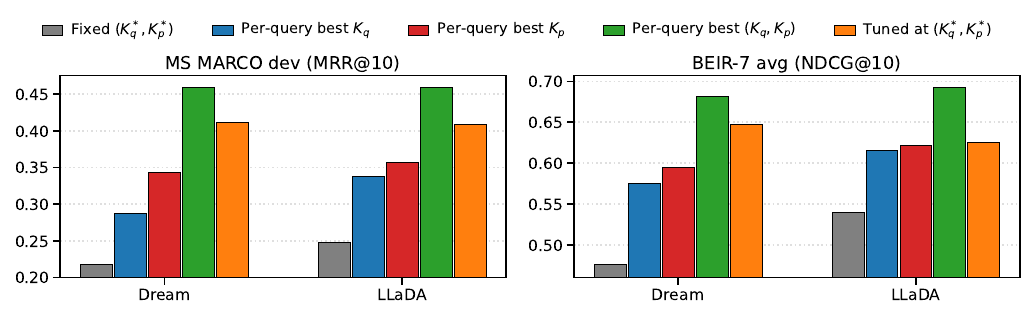}
	\caption{Per-query oracle headroom on MS~MARCO dev (MRR@10)
		and BEIR-7 average (NDCG@10), for Dream and LLaDA. Each
		group compares the fixed deployable budget
		$(K_q^{*}, K_p^{*})$ (grey) against three zero-shot
		per-query oracles adapting $K_q$ only (blue), $K_p$ only
		(red), or both jointly (green), and the contrastively
		fine-tuned model at the same fixed budget (orange).
		Oracles are upper bounds, not deployable systems: they
		serve as ceilings for learned budget selection on
		zero-shot \method{}.}
	\label{fig:oracle_bars}
\end{figure}

Figure~\ref{fig:oracle_bars} reports three zero-shot oracles over
the same budget grid used in the main analysis. The $K_q$-only
oracle adapts $K_q$ per query while holding $K_p$ fixed at
$K_p^{*}$; the $K_p$-only oracle does the converse; and the joint
oracle adapts both $K_q$ and $K_p$. These oracles use retrieval
labels to choose the best budget for each query, so they are upper
bounds rather than deployable systems.

The joint oracle gives large gains over the fixed zero-shot budget
and even exceeds fixed-budget contrastive fine-tuning. On MS~MARCO
dev, it exceeds the fine-tuned fixed-budget model by $.05$ MRR@10
for both Dream and LLaDA. On BEIR-7, it exceeds fine-tuning by $.03$
NDCG@10 for Dream and $.07$ for LLaDA. Since the oracle uses the
frozen zero-shot backbone, the improvement comes from choosing a
better budget for each query, not from changing model parameters.

Not all oracle headroom is equally actionable. Query-side adaptation
is the most practical case because the query is encoded online, so a
predictor can choose $K_q$ before running the model. Passage-side
adaptation is harder because passages are indexed offline: a
per-query choice of $K_p$ would require separate passage indexes for
each candidate $K_p$, while a more deployable alternative would choose
$K_p$ per passage at indexing time. Our current fine-tuning fixes
$(K_q,K_p)$ for all examples, so a practical adaptive system would
also need training under variable budgets.

\begin{figure}[t]
	\centering
	\includegraphics[width=\textwidth]{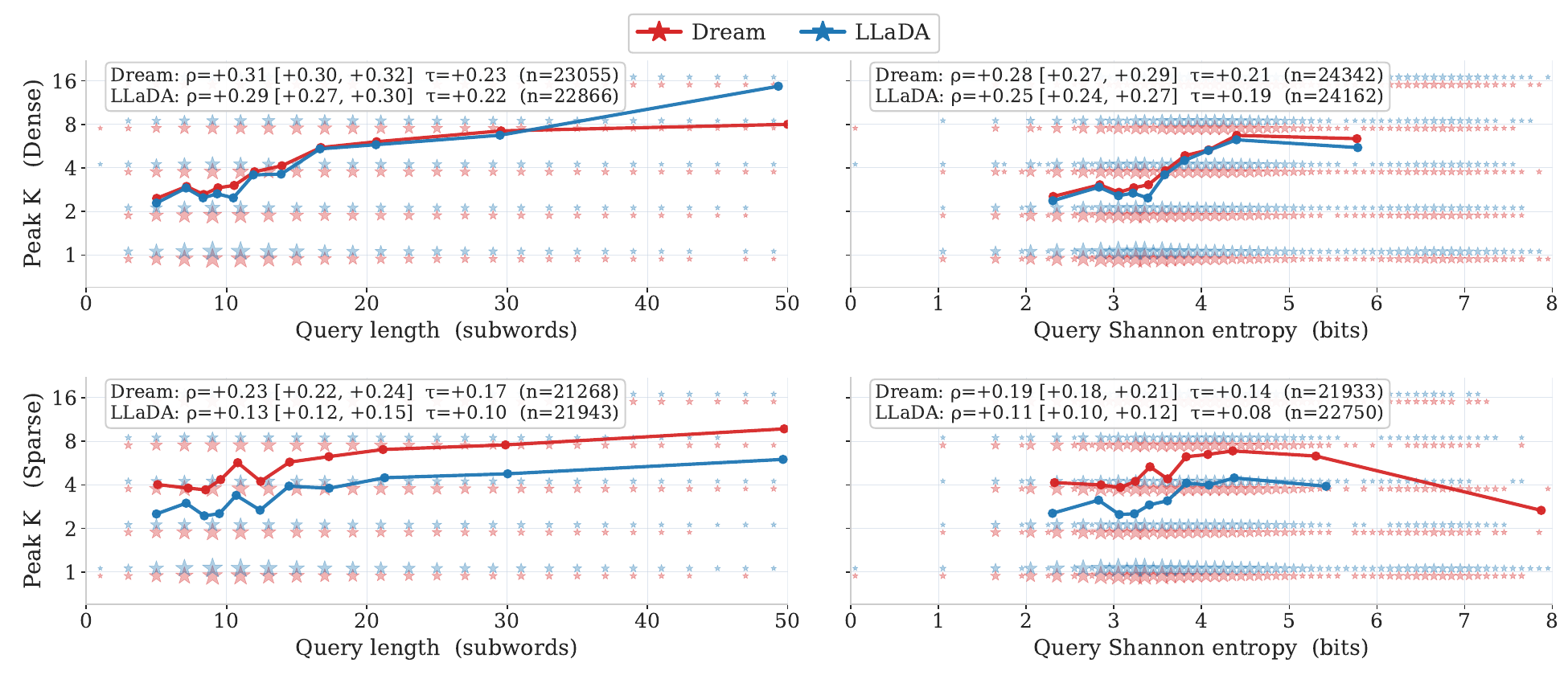}
	\caption{Per-query peak $K_q^{\star}$ vs.\ two cheap query
		features, on Dream and LLaDA. Top row: dense scoring.
		Bottom row: sparse scoring. Left column: query length
		(model-tokenizer subwords). Right column: query Shannon
		entropy (bits, over tokenizer ids). Spearman $\rho$ and
		Kendall $\tau$ are shown in each panel inset, with $95\%$
		bootstrap confidence intervals. Both features correlate
		positively with peak $K_q$ on both backbones; length is
		the stronger signal, and dense scoring shows cleaner
		correlations than sparse.}
	\label{fig:k_query_features}
\end{figure}

We next ask whether the query-side oracle preference is structured
enough to predict before encoding. Figure~\ref{fig:k_query_features}
correlates the oracle-optimal query budget $K_q^{\star}$ with two
cheap query features: query length and query Shannon entropy. For
each query in the pooled MS~MARCO dev, DL19, DL20, and BEIR-7
evaluation sets, $K_q^{\star}$ is the query budget that maximizes
the per-query retrieval metric under the corresponding scoring mode.
Query length is the model-tokenizer subword count, and query entropy
is the Shannon entropy of the query token distribution. We report
Spearman $\rho$ and Kendall $\tau$ across the pooled set.\footnote{The
	query-length analysis pools $9$ datasets, excluding one dataset
	whose query field uses a different preprocessing format; the
	entropy analysis pools the full BEIR-7 plus in-domain evaluation
	sets ($10$ datasets total).}

Both features correlate positively with the oracle-preferred query
budget. Query length is the stronger signal: under dense scoring, it
reaches $\rho{=}+0.31$ for Dream and $\rho{=}+0.29$ for LLaDA, while
under sparse scoring it drops to $\rho{=}+0.23$ and $\rho{=}+0.13$.
Query entropy shows the same trend but with slightly weaker
correlations: $\rho{=}+0.28$ and $\rho{=}+0.25$ under dense scoring,
and $\rho{=}+0.19$ and $\rho{=}+0.11$ under sparse scoring. On the
dense panels, peak $K_q^{\star}$ rises monotonically from
approximately $2$ for the shortest queries to approximately $8$ for
the longest. The sparse relationship is flatter, especially for
LLaDA, suggesting that sparse signals saturate at smaller query
budgets.

These correlations are modest, so length and entropy alone would not
recover the full oracle headroom. They nevertheless show that oracle
budget preferences are partially structured rather than pure noise.
Since both features are available before encoding, they provide a
simple starting point for learned query-side budget selection. A
practical adaptive retriever would need to learn this selector, decide
how to handle passage-side budgets at indexing time, and train the
retriever under variable budgets.

\section{Conclusion}
\label{sec:conclusion}

This paper shows that diffusion language models can be more effective retrievers when retrieval is formulated through their native masked-position prediction interface rather than encoder-style pooling. This interface naturally supports multi-representation retrieval by producing multiple retrieval representations in one forward pass. Even a single masked position
improves over encoder-style diffusion retrieval, while multiple masked
positions further improve effectiveness without the sequential generation cost
that limited prior autoregressive multi-representation retrievers.

Both zero-shot and fine-tuned, \method{} achieves the strongest
aggregate effectiveness within our matched comparison. Our analysis also shows
that different queries benefit from different numbers of masked positions,
motivating adaptive masked-position allocation as a next step.

\section*{Limitations}

Our experiments focus on 7--8B-scale models, comparing two diffusion backbones (Dream and LLaDA) with two autoregressive backbones (Qwen2.5 and LLaMA3). This scale supports a controlled comparison of retrieval interfaces, especially because Dream is initialized from Qwen2.5, giving the closest available pairing between an autoregressive and diffusion backbone. Whether the interface advantage holds at other scales is an open empirical question.

Our evaluation covers MS~MARCO, TREC~DL 2019/2020, and a seven-dataset subset of BEIR, providing in-domain retrieval together with seven out-of-domain transfer tasks across open-domain QA, multi-hop QA, scientific, biomedical, financial, argument, and duplicate-question retrieval. We view this as sufficient to establish the central claim that the masked-position interface improves over encoder-style and autoregressive retrieval at matched scale. Extending the same interface to multilingual or long-document retrieval is a promising direction, which we leave to future work.

\method{} uses fixed query and passage masked-position counts $(K_q,K_p)$, selected once on MS~MARCO training data and reused across all evaluations. These counts transfer well in aggregate, but our per-query oracle analysis (\S\ref{sec:analysis_oracle}) shows substantial headroom from adaptive selection, and the correlations with cheap query features such as length and entropy suggest that a lightweight learned predictor for $(K_q,K_p)$ is a practical and promising avenue.

Multi-representation retrieval also increases passage-index storage, roughly proportional to $K_p$. This is the main systems tradeoff: \method{} avoids the sequential cost of autoregressive query decoding, but still pays a storage cost on the passage side. The cost is much smaller than token-level late-interaction systems like ColBERT, and standard ColBERTv2  compression reduces it by roughly $6\times$ in our preliminary check (App.~\ref{app:index_storage}) with limited effectiveness loss. Compression-aware training and learned per-passage budget allocation are natural ways to push this tradeoff further, which we leave to future work.

\bibliographystyle{plainnat}
\bibliography{custom}

\appendix


\section{Supplementary Method Details}
\label{app:method}

This section provides supplementary method details that support
the design choices in \S\ref{sec:method}.

\subsection{Multi-Step Denoising}
\label{app:multistep}

The main text uses a single forward pass (\S\ref{sec:method_schedule}, $S{=}1$) to process all appended masked positions, since retrieval requires only the masked-position representations rather than committed token predictions. For completeness, this appendix also evaluates the multi-step variant ($S{>}1$), where the model iteratively denoises the masked sequence over $S$ steps by revealing a subset of positions at each step and re-encoding the remaining ones, at the multi-representation budgets used in the main text.

\paragraph{Procedure.}
With $S$ denoising steps, the $K$ masked positions are unmasked
over $S$ rounds following the confidence-based schedule of the
underlying diffusion backbone~\citep{nie2025large,ye2025dream},
with roughly $K/S$ positions unmasked per step. At each step, the
model performs one forward pass and replaces the most confident
remaining \mask{} symbols with their argmax predicted tokens. The
hidden state and logits at each newly unmasked position are stored
in a frozen buffer. At the final step, representations from the
buffer and the current forward pass are combined, and the dense and
sparse scores from \S\ref{sec:method_scoring} are computed over this
combined representation set.

\paragraph{Fine-tuning interaction.}
When multi-step decoding is applied during supervised fine-tuning
(\S\ref{sec:method_finetune}), the retrieval loss is computed only
at the final step. The frozen buffer carries no gradient, while
positions represented in the current final forward pass contribute
gradients normally.

\paragraph{Empirical effect.}
We compare $S{=}1$ and $S{=}2$ for both diffusion backbones at the
train-selected $(K_q^{*}, K_p^{*})$ used in the main text, across
zero-shot and fine-tuned settings on MS~MARCO, DL19, DL20, and the
seven BEIR-7 datasets. Tables~\ref{tab:multistep_indomain}
and~\ref{tab:multistep_beir7} report the full breakdown.

\begin{table*}[t]
	\centering
		\caption{Multi-step denoising on in-domain benchmarks at the train-selected $(K_q^{*}, K_p^{*})$: Dream $(4, 16)$, LLaDA $(4, 4)$. $S{=}1$ is the single-pass variant used in the main text; $S{=}2$ is two-step iterative denoising. The $S$ column is the number of denoising passes; D/S/H denote dense, sparse, and hybrid scores. MS~MARCO dev uses MRR@10; DL19 and DL20 use NDCG@10. \textbf{Bold}: better of the $S{=}1$/$S{=}2$ pair within each (backbone, regime, column). \underline{Underline}: best score across all rows in the column. Paired two-sided t-tests compare $S{=}2$ against $S{=}1$: $\dagger$ ($p<0.05$), $\ddagger$ ($p<0.01$). $S{=}2$ does not consistently improve over $S{=}1$ in any configuration.}
	\label{tab:multistep_indomain}
	\small
	\setlength{\tabcolsep}{3pt}
	\begin{tabular*}{\textwidth}{@{\extracolsep{\fill}}lllccccccccc}
		\toprule
		& & & \multicolumn{3}{c}{MS~MARCO dev} & \multicolumn{3}{c}{DL19} & \multicolumn{3}{c}{DL20} \\
		\cmidrule(lr){4-6} \cmidrule(lr){7-9} \cmidrule(lr){10-12}
		& Regime & $S$ & D & S & H & D & S & H & D & S & H \\
		\midrule
		\multirow{4}{*}{\rotatebox[origin=c]{90}{\textbf{Dream}}} & \multirow{2}{*}{Zero-shot} & $S{=}1$ & $\mathbf{.192}$ & $\mathbf{.165}$ & $\mathbf{.218}$ & $\mathbf{.479}$ & $\mathbf{.427}$ & $\mathbf{.496}$ & $\mathbf{.466}$ & $\mathbf{.422}$ & $\mathbf{.512}$ \\
		&  & $S{=}2$ & $.102^{\ddagger}$ & $.132^{\ddagger}$ & $.172^{\ddagger}$ & $.287^{\ddagger}$ & $.393$ & $.465$ & $.317^{\ddagger}$ & $.373^{\dagger}$ & $.446^{\ddagger}$ \\
		\cmidrule(lr){2-12}
		& \multirow{2}{*}{Fine-tuned} & $S{=}1$ & \underline{$\mathbf{.433}$} & \underline{$\mathbf{.349}$} & \underline{$\mathbf{.411}$} & $.751$ & $.620$ & $.739$ & $.729$ & $.617$ & $.697$ \\
		&  & $S{=}2$ & $.425^{\ddagger}$ & $.344^{\dagger}$ & $.409$ & \underline{$\mathbf{.759}$} & $\mathbf{.622}$ & \underline{$\mathbf{.741}$} & \underline{$\mathbf{.746}$} & \underline{$\mathbf{.625}$} & \underline{$\mathbf{.709}$} \\
		\midrule
		\multirow{4}{*}{\rotatebox[origin=c]{90}{\textbf{LLaDA}}} & \multirow{2}{*}{Zero-shot} & $S{=}1$ & $\mathbf{.192}$ & $\mathbf{.223}$ & $\mathbf{.248}$ & $\mathbf{.490}$ & $\mathbf{.479}$ & $\mathbf{.549}$ & $\mathbf{.472}$ & $\mathbf{.479}$ & $\mathbf{.536}$ \\
		&  & $S{=}2$ & $.132^{\ddagger}$ & $.208^{\ddagger}$ & $.205^{\ddagger}$ & $.392^{\ddagger}$ & $.477$ & $.513$ & $.374^{\ddagger}$ & $.464$ & $.483^{\dagger}$ \\
		\cmidrule(lr){2-12}
		& \multirow{2}{*}{Fine-tuned} & $S{=}1$ & $\mathbf{.427}$ & $\mathbf{.348}$ & $\mathbf{.408}$ & $.718$ & \underline{$\mathbf{.636}$} & $\mathbf{.718}$ & $.721$ & $\mathbf{.614}$ & $\mathbf{.698}$ \\
		&  & $S{=}2$ & $.414^{\ddagger}$ & $.330^{\ddagger}$ & $.397^{\ddagger}$ & $\mathbf{.719}$ & $.624$ & $.715$ & $\mathbf{.724}$ & $.594$ & $.690$ \\
		\bottomrule
	\end{tabular*}
\end{table*}

\begin{table*}[t]
	\centering
	\caption{Multi-step denoising on BEIR-7 (out-of-domain). Per-dataset NDCG@10 plus the BEIR-7 average, at the train-selected $(K_q^{*}, K_p^{*})$: Dream $(4, 16)$, LLaDA $(4, 4)$. $S{=}1$ is the single-pass variant used in the main text; $S{=}2$ is two-step iterative denoising. The $S$ column is the number of denoising passes; D/S/H denote dense, sparse, and hybrid scores. \textbf{Bold}: better of the $S{=}1$/$S{=}2$ pair within each (backbone, regime, column). \underline{Underline}: best score across all rows in the column. Paired two-sided t-tests compare $S{=}2$ against $S{=}1$ (pooled across the seven BEIR datasets for Avg): $\dagger$ ($p<0.05$), $\ddagger$ ($p<0.01$). $S{=}2$ does not consistently improve over $S{=}1$ in any configuration.}
	\label{tab:multistep_beir7}
	\small
	\setlength{\tabcolsep}{3pt}
	\begin{tabular*}{\textwidth}{@{\extracolsep{\fill}}lllcccccccccccc}
		\toprule
		& & & \multicolumn{3}{c}{NQ} & \multicolumn{3}{c}{HQA} & \multicolumn{3}{c}{SciFact} & \multicolumn{3}{c}{COVID} \\
		\cmidrule(lr){4-6} \cmidrule(lr){7-9} \cmidrule(lr){10-12} \cmidrule(lr){13-15}
		& Regime & $S$ & D & S & H & D & S & H & D & S & H & D & S & H \\
		\midrule
		\multirow{4}{*}{\rotatebox[origin=c]{90}{\textbf{Dream}}} & \multirow{2}{*}{Zero-shot} & $S{=}1$ & $\mathbf{.346}$ & $\mathbf{.219}$ & $\mathbf{.358}$ & $\mathbf{.294}$ & $\mathbf{.411}$ & $\mathbf{.449}$ & $\mathbf{.554}$ & $\mathbf{.564}$ & $\mathbf{.636}$ & $\mathbf{.622}$ & $\mathbf{.481}$ & $\mathbf{.590}$ \\
		&  & $S{=}2$ & $.179^{\ddagger}$ & $.173^{\ddagger}$ & $.267^{\ddagger}$ & $.121^{\ddagger}$ & $.292^{\ddagger}$ & $.301^{\ddagger}$ & $.419^{\ddagger}$ & $.461^{\ddagger}$ & $.559^{\ddagger}$ & $.382^{\ddagger}$ & $.313^{\ddagger}$ & $.495^{\ddagger}$ \\
		\cmidrule(lr){2-15}
		& \multirow{2}{*}{Fine-tuned} & $S{=}1$ & \underline{$\mathbf{.644}$} & \underline{$\mathbf{.458}$} & \underline{$\mathbf{.596}$} & \underline{$\mathbf{.683}$} & $.603$ & \underline{$\mathbf{.705}$} & $.752$ & $\mathbf{.666}$ & $\mathbf{.729}$ & \underline{$\mathbf{.847}$} & $\mathbf{.665}$ & \underline{$\mathbf{.830}$} \\
		&  & $S{=}2$ & $.643$ & $.455$ & $.596$ & $.662^{\ddagger}$ & $\mathbf{.608}^{\ddagger}$ & $.697^{\ddagger}$ & $\mathbf{.753}$ & $.666$ & $.723$ & $.825$ & $.658$ & $.789^{\ddagger}$ \\
		\midrule
		\multirow{4}{*}{\rotatebox[origin=c]{90}{\textbf{LLaDA}}} & \multirow{2}{*}{Zero-shot} & $S{=}1$ & $\mathbf{.385}$ & $\mathbf{.292}$ & $\mathbf{.415}$ & $\mathbf{.328}$ & $\mathbf{.468}$ & $\mathbf{.494}$ & $\mathbf{.592}$ & $\mathbf{.624}$ & $\mathbf{.660}$ & $\mathbf{.662}$ & $\mathbf{.690}$ & $\mathbf{.717}$ \\
		&  & $S{=}2$ & $.255^{\ddagger}$ & $.267^{\ddagger}$ & $.328^{\ddagger}$ & $.177^{\ddagger}$ & $.424^{\ddagger}$ & $.385^{\ddagger}$ & $.463^{\ddagger}$ & $.605$ & $.607^{\ddagger}$ & $.571^{\ddagger}$ & $.638^{\ddagger}$ & $.697$ \\
		\cmidrule(lr){2-15}
		& \multirow{2}{*}{Fine-tuned} & $S{=}1$ & $.622$ & $\mathbf{.452}$ & $\mathbf{.584}$ & $.647$ & $.613$ & $.687$ & $.744$ & \underline{$\mathbf{.695}$} & $.746$ & $\mathbf{.846}$ & \underline{$\mathbf{.710}$} & $\mathbf{.819}$ \\
		&  & $S{=}2$ & $\mathbf{.634}^{\ddagger}$ & $.429^{\ddagger}$ & $.582$ & $\mathbf{.660}^{\ddagger}$ & \underline{$\mathbf{.615}$} & $\mathbf{.696}^{\ddagger}$ & \underline{$\mathbf{.761}$} & $.676$ & \underline{$\mathbf{.747}$} & $.841$ & $.659^{\ddagger}$ & $.808$ \\
		\bottomrule
	\end{tabular*}
	\vspace{0.6em}
	\begin{tabular*}{\textwidth}{@{\extracolsep{\fill}}lllcccccccccccc}
		\toprule
		& & & \multicolumn{3}{c}{FiQA} & \multicolumn{3}{c}{ArguAna} & \multicolumn{3}{c}{Quora} & \multicolumn{3}{c}{Avg} \\
		\cmidrule(lr){4-6} \cmidrule(lr){7-9} \cmidrule(lr){10-12} \cmidrule(lr){13-15}
		& Regime & $S$ & D & S & H & D & S & H & D & S & H & D & S & H \\
		\midrule
		\multirow{4}{*}{\rotatebox[origin=c]{90}{\textbf{Dream}}} & \multirow{2}{*}{Zero-shot} & $S{=}1$ & $\mathbf{.293}$ & $\mathbf{.173}$ & $\mathbf{.278}$ & $\mathbf{.350}$ & $\mathbf{.137}$ & $\mathbf{.323}$ & $.532$ & $\mathbf{.647}$ & $.696$ & $\mathbf{.427}$ & $\mathbf{.376}$ & $\mathbf{.476}$ \\
		&  & $S{=}2$ & $.179^{\ddagger}$ & $.127^{\ddagger}$ & $.219^{\ddagger}$ & $.229^{\ddagger}$ & $.080^{\ddagger}$ & $.196^{\ddagger}$ & $\mathbf{.654}^{\ddagger}$ & $.517^{\ddagger}$ & $\mathbf{.707}^{\ddagger}$ & $.309^{\ddagger}$ & $.280^{\ddagger}$ & $.392^{\ddagger}$ \\
		\cmidrule(lr){2-15}
		& \multirow{2}{*}{Fine-tuned} & $S{=}1$ & $.479$ & \underline{$\mathbf{.316}$} & \underline{$\mathbf{.431}$} & $.403$ & \underline{$\mathbf{.305}$} & $.382$ & \underline{$\mathbf{.887}$} & \underline{$\mathbf{.748}$} & \underline{$\mathbf{.859}$} & \underline{$\mathbf{.671}$} & \underline{$\mathbf{.537}$} & \underline{$\mathbf{.647}$} \\
		&  & $S{=}2$ & \underline{$\mathbf{.485}$} & $.313$ & $.429$ & $\mathbf{.417}^{\ddagger}$ & $.295^{\ddagger}$ & $\mathbf{.396}^{\ddagger}$ & $.878^{\ddagger}$ & $.747$ & $.854^{\ddagger}$ & $.666^{\ddagger}$ & $.535$ & $.640^{\ddagger}$ \\
		\midrule
		\multirow{4}{*}{\rotatebox[origin=c]{90}{\textbf{LLaDA}}} & \multirow{2}{*}{Zero-shot} & $S{=}1$ & $\mathbf{.308}$ & $\mathbf{.225}$ & $\mathbf{.317}$ & $\mathbf{.386}$ & $\mathbf{.179}$ & $\mathbf{.357}$ & $\mathbf{.819}$ & $\mathbf{.678}$ & $\mathbf{.814}$ & $\mathbf{.497}$ & $\mathbf{.451}$ & $\mathbf{.539}$ \\
		&  & $S{=}2$ & $.224^{\ddagger}$ & $.215^{\dagger}$ & $.279^{\ddagger}$ & $.324^{\ddagger}$ & $.173^{\dagger}$ & $.309^{\ddagger}$ & $.680^{\ddagger}$ & $.663^{\ddagger}$ & $.754^{\ddagger}$ & $.385^{\ddagger}$ & $.427^{\ddagger}$ & $.480^{\ddagger}$ \\
		\cmidrule(lr){2-15}
		& \multirow{2}{*}{Fine-tuned} & $S{=}1$ & $.443$ & $\mathbf{.303}$ & $.397$ & $.412$ & $.291$ & $.389$ & $.798$ & $.595$ & $.760$ & $.645$ & $\mathbf{.523}$ & $.626$ \\
		&  & $S{=}2$ & $\mathbf{.465}^{\ddagger}$ & $.294$ & $\mathbf{.412}^{\ddagger}$ & \underline{$\mathbf{.428}^{\ddagger}$} & $\mathbf{.303}^{\ddagger}$ & \underline{$\mathbf{.406}^{\ddagger}$} & $\mathbf{.842}^{\ddagger}$ & $\mathbf{.654}^{\ddagger}$ & $\mathbf{.803}^{\ddagger}$ & $\mathbf{.662}^{\ddagger}$ & $.519^{\ddagger}$ & $\mathbf{.636}^{\ddagger}$ \\
		\bottomrule
	\end{tabular*}
\end{table*}

At zero-shot, $S{=}1$ wins decisively. $S{=}2$ is significantly
worse than $S{=}1$ on every benchmark and every BEIR-7 dataset for
both backbones, often by large margins: Dream MS~MARCO hybrid drops
from $.218$ to $.172$ ($\ddagger$), and the BEIR-7 dense average
drops from $.427$ to $.309$ for Dream and from $.497$ to $.385$ for
LLaDA (both $\ddagger$). These results support the single-pass choice
used in the main text: zero-shot retrieval benefits from processing
the appended masked positions directly, while iterative denoising
degrades effectiveness in this setting.

After fine-tuning, the effect depends on the backbone. For Dream,
$S{=}2$ has no consistent benefit: MS~MARCO dense and BEIR-7 dense
average both decrease, with only ArguAna showing a significant BEIR-7
gain. For LLaDA, $S{=}2$ improves more broadly, raising the BEIR-7
dense average from $.645$ to $.662$ ($\ddagger$), but MS~MARCO dense
drops from $.427$ to $.414$. Because $S{=}1$ is consistently best at
zero-shot, avoids backbone-specific inference choices, and preserves
the single-pass latency advantage, we use $S{=}1$ throughout the main
text. Selective $S{>}1$ decoding for specific backbones or domains is
left to future work.

\subsection{Prompt Comparison: ``A Few Words'' vs.\ ``Three
	Words''}
\label{app:prompts}

\begin{table*}[t]
	\centering
	\caption{Prompt phrasing ablation for zero-shot PromptReps multi-rep retrieval: MRR@10 for MS~MARCO dev and NDCG@10 for TREC DL. D/S/H denote dense, sparse, and hybrid scores. \textbf{Bold}: better phrasing for each backbone and column, including rounded ties. Paired two-sided t-tests compare the two phrasings; significant pairs carry $\dagger$ ($p<0.05$) or $\ddagger$ ($p<0.01$).}
	\label{tab:prompt_comparison}
	\small
	\setlength{\tabcolsep}{4pt}
	\begin{tabular*}{\textwidth}{@{\extracolsep{\fill}}llccccccccc}
		\toprule
		& & \multicolumn{3}{c}{MS~MARCO dev} & \multicolumn{3}{c}{DL19} & \multicolumn{3}{c}{DL20} \\
		\cmidrule(lr){3-5} \cmidrule(lr){6-8} \cmidrule(lr){9-11}
		Backbone & Phrasing & D & S & H & D & S & H & D & S & H \\
		\midrule
		\multirow{2}{*}{LLaMA3-8B} & ``three words'' & $.094^{\ddagger}$ & $.194$ & $.181^{\ddagger}$ & $.216^{\ddagger}$ & $.416$ & $.388^{\ddagger}$ & $.203^{\ddagger}$ & $\mathbf{.416}$ & $.407^{\ddagger}$ \\
		& ``a few words'' & $\mathbf{.151}^{\ddagger}$ & $\mathbf{.196}$ & $\mathbf{.220}^{\ddagger}$ & $\mathbf{.319}^{\ddagger}$ & $\mathbf{.454}$ & $\mathbf{.473}^{\ddagger}$ & $\mathbf{.336}^{\ddagger}$ & $.412$ & $\mathbf{.471}^{\ddagger}$ \\
		\cmidrule(lr){2-11}
		\multirow{2}{*}{Qwen2.5-7B} & ``three words'' & $.142$ & $\mathbf{.200}^{\ddagger}$ & $\mathbf{.211}^{\ddagger}$ & $.383$ & $\mathbf{.445}^{\dagger}$ & $\mathbf{.485}$ & $\mathbf{.390}$ & $\mathbf{.436}^{\dagger}$ & $\mathbf{.495}$ \\
		& ``a few words'' & $.142$ & $.177^{\ddagger}$ & $.200^{\ddagger}$ & $\mathbf{.404}$ & $.405^{\dagger}$ & $.477$ & $.386$ & $.410^{\dagger}$ & $.478$ \\
		\bottomrule
	\end{tabular*}
\end{table*}

\promptreps{}'s multi-token representation variant~\citep{zhuang2024promptreps} instructs the model to produce \textit{three words} as a short textual representation of the input. We use \textit{a few words} instead (Table~\ref{tab:prompt}), since the variable-budget setting in \method{} is not tied to a fixed count of three and the more flexible phrasing aligns better with sweeping over different $K$ values. Since the original \promptreps{} multi-representation experiments were conducted on LLaMA3-8B, we compare both phrasings on LLaMA3-8B and Qwen2.5-7B.
This check follows evidence that prompt wording can affect zero-shot
LLM retrieval and ranking behavior~\citep{sun2025investigation}.

\paragraph{Setup.}
We run zero-shot retrieval with multiple generated representations on
MS~MARCO dev, TREC~DL19, and TREC~DL20 under each phrasing, holding
all other settings fixed: the same scoring functions, hybrid
interpolation, and $N{=}20$ zero-shot cap on autoregressive sequential
generation.

\paragraph{Result.}
Table~\ref{tab:prompt_comparison} reports the comparison, and the two
backbones differ. On LLaMA3-8B, \textit{``a few words''} wins on $8$
of $9$ benchmark--scoring cells, with significant dense and hybrid
gains across all three benchmarks. For example, dense MRR@10 on
MS~MARCO rises from $.094$ to $.151$, and dense NDCG@10 on DL19 rises
from $.216$ to $.319$. On Qwen2.5-7B, the ranking is less uniform:
\textit{``three words''} wins $5$ of $9$ cells, mostly on sparse and
hybrid, while dense is statistically tied on MS~MARCO and only
marginally different on DL19 and DL20. Thus, the two phrasings are not
strictly ordered across backbones; the gap is larger for LLaMA3 than
for Qwen2.5, and the win direction differs between them.

We use \textit{``a few words''} as the default phrasing in the main
text for three reasons. First, it does not impose a fixed semantic
count such as \textit{``three''}, which is better aligned with a
variable-representation setting. Second, it improves the original
\promptreps{} backbone, LLaMA3, where the gap between phrasings is
largest. Third, the same prompt is used uniformly across all backbones,
so comparisons between decoding strategies are not affected by
backbone-specific prompt tuning.


\section{Supplementary Experimental Setup}
\label{app:exp}

This section gives implementation and measurement details for the
experiments in \S\ref{sec:experimental_setup}.

\subsection{Fine-Tuning Details}
\label{app:finetune_details}

Table~\ref{tab:finetune_hparams} lists the fine-tuning recipe.
We use the same optimization, batching, and LoRA settings across
all four backbones and re-trained baselines, so fine-tuned
comparisons are not confounded by different training recipes.
LoRA target module names differ between LLaMA-/Qwen-style
backbones and LLaDA, but we target the same functional set of
weight matrices: attention $\{q,k,v,o\}$ projections and the
three feed-forward projections. The trainable-parameter fraction
stays within $0.52$--$0.53\%$ across all four backbones
($40$--$42$\,M parameters).

\begin{center}
	\centering
	\small
	\setlength{\tabcolsep}{4pt}
	\renewcommand{\arraystretch}{0.94}
	\captionof{table}{Fine-tuning hyperparameters. Applied uniformly to all four backbones and re-trained baselines.}
	\label{tab:finetune_hparams}
	\begin{tabularx}{\textwidth}{@{}lXlX@{}}
		\toprule
		\multicolumn{2}{c}{\textbf{Training}} &
		\multicolumn{2}{c}{\textbf{Batching}} \\
		\midrule
		Optimizer & AdamW & Per-device batch & $8$ \\
		Learning rate & $1{\times}10^{-4}$ & Grad.\ accumulation & $4$ \\
		Warmup ratio & $0.06$ & Global batch & $128$ \\
		Epochs & $1$ & Negatives per query & $1$ positive $+\;15$ hard \\
		Precision & bf16 & Query / passage length & $32$ / $156$ tokens \\
		Distributed / hardware & ZeRO-2, $4{\times}$H100 & InfoNCE temperature & $\tau = 0.01$ \\
		Random seed & 42 & & \\
		GPU hours & 10--15 hours per model & & \\
		\midrule
		\multicolumn{2}{c}{\textbf{LoRA}} &
		\multicolumn{2}{c}{\textbf{Method-specific}} \\
		\midrule
		Rank $r$ & $16$ & Diffusion $S$ at training & $1$ \\
		Scaling $\alpha$ & $64$ & Dream $(K_q, K_p)$ & $(4, 16)$ \\
		Dropout & $0.05$ & LLaDA $(K_q, K_p)$ & $(4, 4)$ \\
		Trainable params & $40$--$42$\,M & \promptreps{} generation (train) & $N{=}4$ \\
		Fraction & $0.52$--$0.53\%$ & \promptreps{} generation (zero-shot) & $N{=}20$ \\
		& & Decoding temperature & $0.0$ \\
		\midrule
		\multicolumn{4}{c}{\textbf{Model Backbone Checkpoint}} \\
		\midrule
		Dream & \multicolumn{3}{l}{Dream-org/Dream-v0-Instruct-7B} \\
		LLaDA & \multicolumn{3}{l}{GSAI-ML/LLaDA-8B-Instruct} \\
		LLaMA3 & \multicolumn{3}{l}{meta-llama/Meta-Llama-3-8B-Instruct} \\
		Qwen2.5 & \multicolumn{3}{l}{Qwen/Qwen2.5-7B-Instruct} \\
		\bottomrule
	\end{tabularx}
\end{center}

The autoregressive baselines train with a maximum generation
budget of $N{=}4$, rather than the zero-shot cap $N{=}20$,
because longer sequential generation exceeds our memory budget
at the global batch size of $128$ (see \S\ref{sec:exp_finetune}).

\subsection{Latency Scaling}
\label{app:latency_setup}

Figure~\ref{fig:teaser} reports per-query latency
at one operating point: the zero-shot autoregressive
multi-representation cap $N{=}20$ on a $100$K-document MS~MARCO
sample. This subsection extends the measurement across two axes:
encoding latency as a function of input sequence length, and
search latency as a function of index size. Measurements use a
single H100 ($96$GB) GPU with the same attention implementation
across all backbones. Runs use synthetic random-token inputs and
random-vector indices to isolate systems cost from retrieval
correctness.

\begin{figure*}[t]
	\centering
	\includegraphics[width=\textwidth]{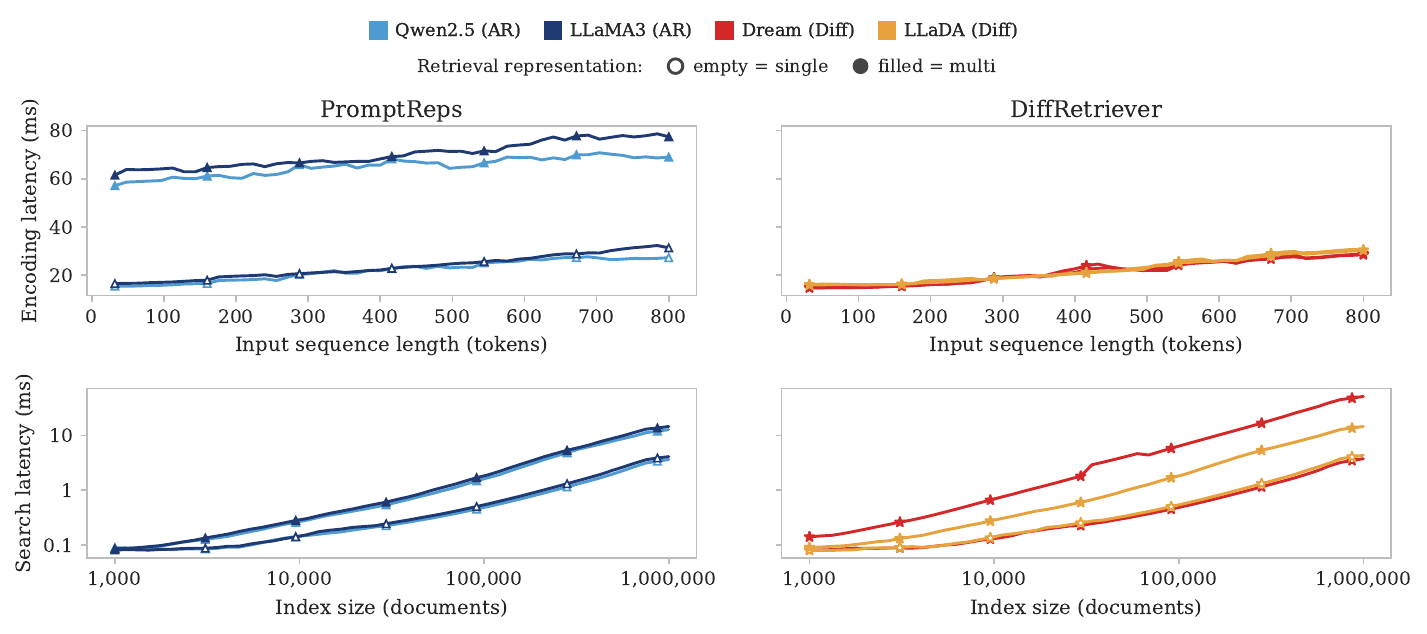}
	\caption{Latency scaling on synthetic inputs and indices.
		Top row: encoding latency vs.\ input sequence length
		(tokens). Bottom row: search latency vs.\ index size
		(documents, log scale). Left column: \promptreps{} on
		autoregressive backbones (Qwen2.5, LLaMA3). Right
		column: \method{} on diffusion backbones (Dream, LLaDA).
		Open markers: single-representation. Filled markers:
		multi-representation (autoregressive models use
		$N{=}4$, the fine-tuned cap; diffusion models use the
		train-selected $(K_q^{*}, K_p^{*})$). All measurements
		use a single H100 GPU with the same attention
		implementation across backbones, $5$ warmup queries, and
		$20$ timed runs averaged.}
	\label{fig:latency_combined}
\end{figure*}

Figure~\ref{fig:latency_combined} reports both axes for the four
main backbones, in two retrieval interfaces (\promptreps{}
and \method{}), at single- and multi-representation configurations.
The multi-representation configurations use the fine-tuned cap
$N{=}4$ for autoregressive backbones (\S\ref{sec:exp_finetune})
and the train-selected $(K_q^{*}, K_p^{*})$ for diffusion backbones
(Dream $(4,16)$, LLaDA $(4,4)$).

\paragraph{Encoding scales mildly with input length on both
	families.}
On the encoding axis (top row of Figure~\ref{fig:latency_combined}),
all systems scale roughly linearly with input length, but at
different absolute levels. Autoregressive multi-representation
retrieval sits at $60$ to $80$~ms across the range;
autoregressive single-representation retrieval sits at $15$ to
$30$~ms. \method{} with multiple representations and \method{}
with one representation both sit near $15$ to $30$~ms, with
multi-representation overhead of roughly $5$ to $10$~ms. At
$N{=}4$, the autoregressive-vs.-diffusion encoding ratio is
approximately $3\times$ at short inputs and narrows toward longer
inputs. The larger ratio in Figure~\ref{fig:teaser} ($\approx15\times$)
comes from the zero-shot $N{=}20$ cap, where sequential decoding
cost compounds.

\paragraph{Search cost grows with index size and passage budget.}
On the search axis (bottom row of Figure~\ref{fig:latency_combined}),
all systems scale predictably with index size under the same
search implementation. Multi-representation retrievers have higher
search cost than single-representation retrievers because each query
performs $K_q$ lookups and the index stores $K_p$ vectors per
document. The difference between \method{} with Dream ($K_p{=}16$)
and \method{} with LLaDA ($K_p{=}4$) is driven by the passage
budget: Dream stores $4\times$ as many vectors per document, which
translates into higher per-query search cost. At $1$M documents,
the spread between the slowest multi-representation configuration
($\approx17$~ms, \method{} with Dream) and the fastest
single-representation configuration ($\approx4$~ms) remains smaller
than the encoding-side gap.

\paragraph{Implication.}
At the fine-tuned autoregressive cap of $N{=}4$, the encoding
advantage of \method{} over \promptreps{} narrows relative to the
zero-shot $N{=}20$ setting, but \method{} multi-representation
encoding remains close to single-representation encoding, while
autoregressive multi-representation encoding remains $2$--$3\times$
slower than autoregressive single-representation encoding. Within
the measured range up to $1$M documents, the encoding-side gap
remains the dominant contributor to end-to-end latency. For larger
deployments, the passage index of multi-representation retrieval,
especially Dream's $K_p{=}16$, may become a more substantial
component of total cost.

\begin{figure*}[t]
	\centering
	\includegraphics[width=\textwidth]{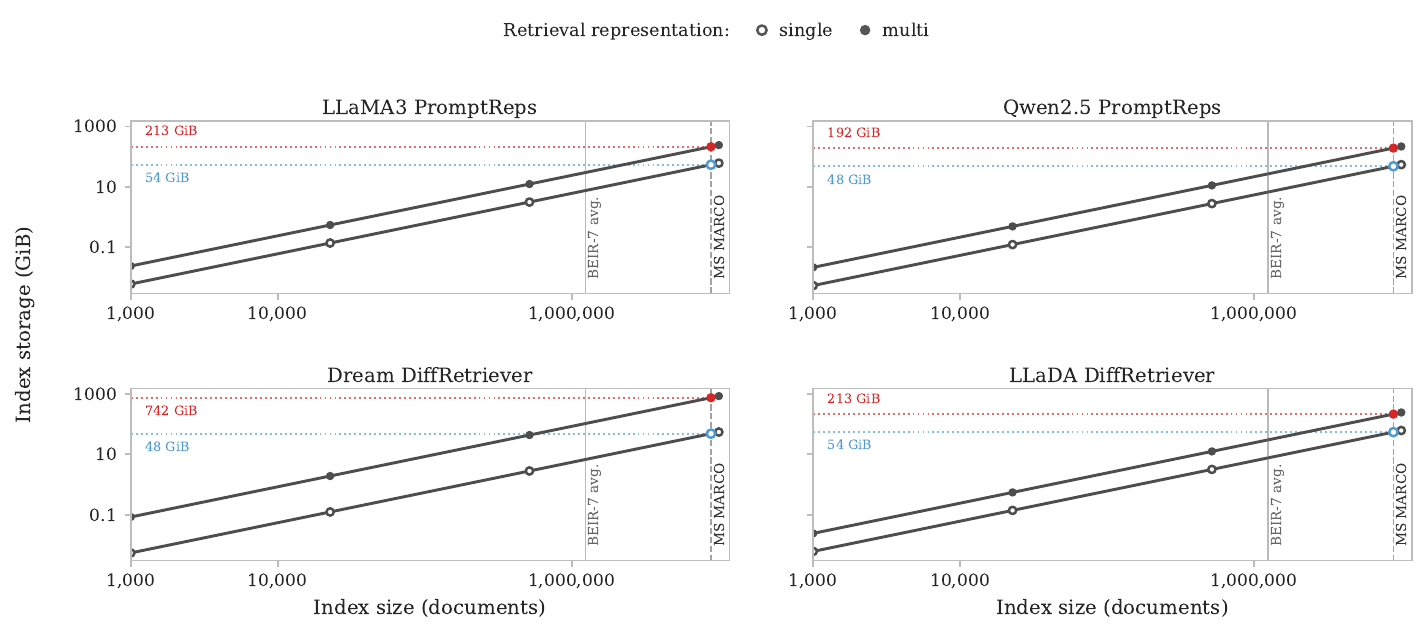}
	\caption{Index storage scaling as a function of corpus size.
		Open markers denote single-representation variants and filled
		markers denote multi-representation variants. Vertical reference
		lines mark the average BEIR-7 corpus size and the MS~MARCO passage
		corpus size. Horizontal dashed lines mark the corresponding
		MS~MARCO-scale storage for single- and multi-representation
		variants. Storage grows approximately linearly with corpus size and
		with the number of passage representations $K_p$.}
	\label{fig:index_storage}
\end{figure*}

\subsection{Index Storage Scaling}
\label{app:index_storage}

Multi-representation retrieval changes the storage footprint of the
passage index because each passage stores $K_p$ retrieval
representations instead of one. We therefore measure index storage as
a function of corpus size.

Figure~\ref{fig:index_storage} shows that storage grows approximately
linearly with the number of indexed documents and with the passage
budget $K_p$. Single-representation systems have the smallest
footprint, while multi-representation systems require proportionally
more storage. This is the main systems tradeoff: \method{} reduces the
sequential encoding cost of producing multiple query representations,
but it does not remove the cost of indexing multiple passage
representations.

At MS~MARCO scale, LLaDA-based \method{} with $(K_q,K_p)=(4,4)$
requires about $213$~GiB for the multi-representation index, compared
with about $54$~GiB for the corresponding single-representation
index. Dream-based \method{} uses a larger passage budget,
$(K_q,K_p)=(4,16)$, and requires about $742$~GiB, compared with about
$47$~GiB for its single-representation variant.

These results clarify the scope of the efficiency claim. The main
advantage of \method{} is that multiple retrieval representations are
produced in one model forward pass, avoiding the sequential decoding
cost of autoregressive multi-representation retrievers. At the same
time, the passage-index storage cost can be substantial when $K_p$ is
large. This cost remains below token-level late-interaction systems
such as ColBERT because \method{} stores a fixed number of
representations per passage rather than one per token.

\paragraph{Reducing the storage cost with compression.}
Standard index-compression techniques from the late-interaction
literature apply directly to \method{} because each passage stores a
fixed number of dense vectors. We follow the ColBERTv2 / PLAID
recipe~\citep{santhanam-etal-2022-colbertv2,santhanam2022plaid}.
First, every indexed vector is decomposed into a centroid, assigned by
$k$-means clustering over all indexed vectors, plus a residual that
captures its offset from that centroid; only the centroid id and the
residual are stored, not the original vector. Second, the residual is
quantized to a small number of bits per dimension (2 bits in our
setup, matching the ColBERTv2 default), which is where most of the
storage saving comes from. Third, at search time, an approximate
nearest-neighbor lookup over the centroids restricts MaxSim scoring to
passages whose vectors lie in the most query-relevant clusters; for
those candidates, vectors are reconstructed from centroid + dequantized
residual and scored against the query as in uncompressed MaxSim. The
result is a compact index that supports the same scoring function used
throughout the paper.

Table~\ref{tab:compressed-storage-effectiveness} reports a preliminary
check on MS~MARCO, TREC~DL, and BEIR-7, applying the same compression
scheme to all single- and multi-representation \method{} indexes.
Across backbones, datasets, and budgets, storage shrinks by roughly
$6\times$ (e.g., Dream multi-representation on MS~MARCO dev:
$742$~GiB~$\rightarrow$~$118$~GiB; LLaDA multi-representation on
MS~MARCO dev: $213$~GiB~$\rightarrow$~$34$~GiB), with modest
effectiveness changes: most paired comparisons differ by no more than
$\sim$0.01--0.03 in the relevant metric, and several BEIR-7 datasets
(DL19, DL20, SciFact) show no significant drop. A small number of
cases, notably LLaDA on Quora, even improve slightly after
compression. These results indicate that the headline storage numbers
above are an upper bound on deployed cost: the multi-representation
index can be compressed by roughly $6\times$ with limited
effectiveness loss, narrowing the practical gap to
single-representation retrieval. We leave a more thorough study of
compression-aware training and indexing to future work.

\begin{center}
	\centering
	\small
	\captionof{table}{Compressed DiffRetriever storage/effectiveness. Eff. and Storage show original$\to$compressed; effectiveness is MRR@10 for MS~MARCO Dev and NDCG@10 otherwise. $^{*}$ marks a paired t-test difference between original and compressed per-query scores ($p<0.05$).}
	\label{tab:compressed-storage-effectiveness}
		\setlength{\tabcolsep}{2.5pt}
	\begin{tabular*}{\columnwidth}{@{\extracolsep{\fill}}lllccc}
		\toprule
		Dataset & Backbone & Variant & Eff. & Storage & Ratio \\
		\midrule
		\multirow{4}{*}{Dev} & Dream & single-rep & .424$\to$.412$^{*}$ & 48G$\to$7.5G & 6.4$\times$ \\
		& Dream & multi-rep & .433$\to$.422$^{*}$ & 742G$\to$118G & 6.3$\times$ \\
		& LLaDA & single-rep & .424$\to$.406$^{*}$ & 54G$\to$8.5G & 6.4$\times$ \\
		& LLaDA & multi-rep & .427$\to$.415$^{*}$ & 213G$\to$34G & 6.3$\times$ \\
		\midrule
		\multirow{4}{*}{DL19} & Dream & single-rep & .741$\to$.729 & 48G$\to$7.5G & 6.4$\times$ \\
		& Dream & multi-rep & .751$\to$.734$^{*}$ & 742G$\to$118G & 6.3$\times$ \\
		& LLaDA & single-rep & .715$\to$.712 & 54G$\to$8.5G & 6.4$\times$ \\
		& LLaDA & multi-rep & .718$\to$.716 & 213G$\to$34G & 6.3$\times$ \\
		\midrule
		\multirow{4}{*}{DL20} & Dream & single-rep & .721$\to$.706 & 48G$\to$7.5G & 6.4$\times$ \\
		& Dream & multi-rep & .729$\to$.722 & 742G$\to$118G & 6.3$\times$ \\
		& LLaDA & single-rep & .715$\to$.712 & 54G$\to$8.5G & 6.4$\times$ \\
		& LLaDA & multi-rep & .721$\to$.732 & 213G$\to$34G & 6.3$\times$ \\
		\midrule
		\multirow{4}{*}{NQ} & Dream & single-rep & .619$\to$.591$^{*}$ & 15G$\to$2.3G & 6.5$\times$ \\
		& Dream & multi-rep & .644$\to$.642 & 225G$\to$36G & 6.3$\times$ \\
		& LLaDA & single-rep & .620$\to$.606$^{*}$ & 17G$\to$2.6G & 6.4$\times$ \\
		& LLaDA & multi-rep & .622$\to$.609$^{*}$ & 65G$\to$11G & 6.3$\times$ \\
		\midrule
		\multirow{4}{*}{HQA} & Dream & single-rep & .650$\to$.622$^{*}$ & 29G$\to$4.4G & 6.4$\times$ \\
		& Dream & multi-rep & .683$\to$.674$^{*}$ & 439G$\to$70G & 6.3$\times$ \\
		& LLaDA & single-rep & .640$\to$.618$^{*}$ & 32G$\to$5.1G & 6.4$\times$ \\
		& LLaDA & multi-rep & .647$\to$.605$^{*}$ & 126G$\to$20G & 6.3$\times$ \\
		\midrule
		\multirow{4}{*}{SciFact} & Dream & single-rep & .739$\to$.729 & 30M$\to$6.3M & 4.7$\times$ \\
		& Dream & multi-rep & .752$\to$.741 & 446M$\to$73M & 6.1$\times$ \\
		& LLaDA & single-rep & .733$\to$.720 & 34M$\to$7.2M & 4.7$\times$ \\
		& LLaDA & multi-rep & .744$\to$.740 & 129M$\to$23M & 5.8$\times$ \\
		\midrule
		\multirow{4}{*}{COVID} & Dream & single-rep & .841$\to$.815$^{*}$ & 962M$\to$150M & 6.5$\times$ \\
		& Dream & multi-rep & .847$\to$.835 & 15G$\to$2.3G & 6.3$\times$ \\
		& LLaDA & single-rep & .840$\to$.811$^{*}$ & 1.1G$\to$171M & 6.4$\times$ \\
		& LLaDA & multi-rep & .846$\to$.822$^{*}$ & 4.2G$\to$663M & 6.4$\times$ \\
		\midrule
		\multirow{4}{*}{FiQA} & Dream & single-rep & .463$\to$.447$^{*}$ & 320M$\to$52M & 6.2$\times$ \\
		& Dream & multi-rep & .479$\to$.469$^{*}$ & 4.9G$\to$790M & 6.3$\times$ \\
		& LLaDA & single-rep & .453$\to$.443$^{*}$ & 364M$\to$59M & 6.2$\times$ \\
		& LLaDA & multi-rep & .443$\to$.437 & 1.4G$\to$228M & 6.2$\times$ \\
		\midrule
		\multirow{4}{*}{ArguAna} & Dream & single-rep & .406$\to$.405 & 49M$\to$9.3M & 5.3$\times$ \\
		& Dream & multi-rep & .403$\to$.393$^{*}$ & 743M$\to$120M & 6.2$\times$ \\
		& LLaDA & single-rep & .414$\to$.400$^{*}$ & 56M$\to$11M & 5.3$\times$ \\
		& LLaDA & multi-rep & .412$\to$.407$^{*}$ & 214M$\to$36M & 6.0$\times$ \\
		\midrule
		\multirow{4}{*}{Quora} & Dream & single-rep & .859$\to$.858 & 2.8G$\to$451M & 6.3$\times$ \\
		& Dream & multi-rep & .887$\to$.885$^{*}$ & 44G$\to$6.7G & 6.6$\times$ \\
		& LLaDA & single-rep & .799$\to$.821$^{*}$ & 3.2G$\to$515M & 6.2$\times$ \\
		& LLaDA & multi-rep & .798$\to$.831$^{*}$ & 13G$\to$2.0G & 6.3$\times$ \\
		\bottomrule
	\end{tabular*}
\end{center}


\section{Supplementary Results}
\label{app:results}

This section reports additional result breakdowns that support
the main comparisons in \S\ref{sec:results}.

\subsection{Full BEIR-7 Breakdown: Dense, Sparse, and
	Hybrid}
\label{app:beir7_full}

\begin{table}[H]
	\centering
	\scriptsize
	\setlength{\tabcolsep}{2pt}
	\caption{BEIR-7 D/S/H results: NDCG@10 dense, sparse, and hybrid scores for every system in Table~\ref{tab:main_beir7}. Fwd denotes query-side forward steps. \textbf{Bold}: best within each half and column, including rounded ties. $\dagger$: better than same-backbone single-rep; $\ddagger$: \method{} multi-rep better than corresponding \promptreps{} single-rep (Dream$\rightarrow$Qwen2.5, LLaDA$\rightarrow$LLaMA3). Paired t-test, $p<0.05$.}
	\label{tab:appendix_beir7_dsh}
	\resizebox{\textwidth}{!}{%
	\begin{tabular}{@{}clllccccccccccccc@{}}
		\toprule
		& \multirow{2}{*}{Method} & \multirow{2}{*}{Backbone} & \multirow{2}{*}{Variant} & \multirow{2}{*}{Fwd} & \multicolumn{3}{c}{NQ} & \multicolumn{3}{c}{HQA} & \multicolumn{3}{c}{SciFact} & \multicolumn{3}{c}{COVID} \\
		\cmidrule(lr){6-8} \cmidrule(lr){9-11} \cmidrule(lr){12-14} \cmidrule(lr){15-17}
		&        &          &         &     & D & S & H & D & S & H & D & S & H & D & S & H \\
		\midrule
		\multirow{11}{*}{\rotatebox[origin=c]{90}{\textbf{Zero-shot}}}
		& BM25                                & ---                            & sparse     & ---      & --- & .243 & --- & --- & \textbf{.567} & --- & --- & \textbf{.664} & --- & --- & .530 & --- \\
		\cmidrule(lr){2-17}
		& \multirow{2}{*}{DiffEmbed}          & Dream                          & single-rep & 1        & .118 & --- & --- & .149 & --- & --- & .383 & --- & --- & .316 & --- & --- \\
		&                                     & LLaDA                          & single-rep & 1        & .066 & --- & --- & .167 & --- & --- & .385 & --- & --- & .407 & --- & --- \\
		\cmidrule(lr){2-17}
		& \multirow{4}{*}{PromptReps}         & \multirow{2}{*}{Qwen2.5}       & single-rep & 1        & .156 & .261 & .302 & .066 & .394 & .322 & .455 & .570 & .618 & .490 & .545 & .577 \\
		&                                     &                                & multi-rep  & $\leq$20 & .232$^\dagger$ & .265 & .316$^\dagger$ & .267$^\dagger$ & .484$^\dagger$ & .470$^\dagger$ & .489 & .595$^\dagger$ & .609 & .411 & .513 & .538 \\
		&                                     & \multirow{2}{*}{LLaMA3}        & single-rep & 1        & .331 & .293 & .410 & .199 & .452 & .454 & .517 & .584 & .643 & .612 & .558 & .645 \\
		&                                     &                                & multi-rep  & $\leq$20 & .301 & .276 & .379 & .287$^\dagger$ & .508$^\dagger$ & \textbf{.507}$^\dagger$ & .539 & .605 & .639 & .530 & .552 & .613 \\
		\cmidrule(lr){2-17}
		& \multirow{4}{*}{\textbf{DiffRetriever}} & \multirow{2}{*}{Dream}         & single-rep & 1        & .117 & .102 & .153 & .038 & .056 & .066 & .334 & .276 & .380 & .501 & .339 & .497 \\
		&                                     &                                & multi-rep  & 1        & .346$^{\dagger\ddagger}$ & .219$^\dagger$ & .358$^{\dagger\ddagger}$ & .294$^{\dagger\ddagger}$ & .411$^{\dagger\ddagger}$ & .449$^{\dagger\ddagger}$ & .554$^{\dagger\ddagger}$ & .564$^\dagger$ & .636$^\dagger$ & .622$^{\dagger\ddagger}$ & .481$^\dagger$ & .590$^\dagger$ \\
		&                                     & \multirow{2}{*}{LLaDA}         & single-rep & 1        & .289 & \textbf{.296} & .382 & .129 & .322 & .299 & .525 & .576 & .618 & \textbf{.679} & .666 & \textbf{.752} \\
		&                                     &                                & multi-rep  & 1        & \textbf{.385}$^{\dagger\ddagger}$ & .292 & \textbf{.415}$^\dagger$ & \textbf{.328}$^{\dagger\ddagger}$ & .468$^{\dagger\ddagger}$ & .494$^{\dagger\ddagger}$ & \textbf{.592}$^{\dagger\ddagger}$ & .624$^{\dagger\ddagger}$ & \textbf{.660}$^\dagger$ & .662 & \textbf{.690}$^\ddagger$ & .717$^\ddagger$ \\
		\midrule
		\multirow{11}{*}{\rotatebox[origin=c]{90}{\textbf{Fine-tuned}}}
		& RepLLaMA                            & LLaMA3                         & single-rep & 1        & .622 & --- & --- & .503 & --- & --- & .653 & --- & --- & .564 & --- & --- \\
		\cmidrule(lr){2-17}
		& \multirow{2}{*}{DiffEmbed}          & Dream                          & single-rep & 1        & .625 & --- & --- & .668 & --- & --- & .735 & --- & --- & .758 & --- & --- \\
		&                                     & LLaDA                          & single-rep & 1        & .587 & --- & --- & .601 & --- & --- & .714 & --- & --- & .578 & --- & --- \\
		\cmidrule(lr){2-17}
		& \multirow{4}{*}{PromptReps}         & \multirow{2}{*}{Qwen2.5}       & single-rep & 1        & .619 & .441 & .573 & .661 & .580 & .684 & .756 & .662 & .727 & .843 & .632 & .797 \\
		&                                     &                                & multi-rep  & $\leq$4  & .623 & .453$^\dagger$ & .582$^\dagger$ & .686$^\dagger$ & .601$^\dagger$ & .701$^\dagger$ & \textbf{.773}$^\dagger$ & .681$^\dagger$ & \textbf{.748}$^\dagger$ & \textbf{.848} & .664 & .818 \\
		&                                     & \multirow{2}{*}{LLaMA3}        & single-rep & 1        & .619 & .443 & .573 & .685 & .595 & .695 & .750 & .659 & .731 & .815 & .596 & .756 \\
		&                                     &                                & multi-rep  & $\leq$4  & .631$^\dagger$ & \textbf{.458}$^\dagger$ & .583$^\dagger$ & \textbf{.696}$^\dagger$ & \textbf{.613}$^\dagger$ & \textbf{.709}$^\dagger$ & .735 & .673 & .737 & .809 & .659$^\dagger$ & .792$^\dagger$ \\
		\cmidrule(lr){2-17}
		& \multirow{4}{*}{\textbf{DiffRetriever}} & \multirow{2}{*}{Dream}         & single-rep & 1        & .619 & .433 & .572 & .650 & .587 & .684 & .739 & .670 & .729 & .841 & .629 & .789 \\
		&                                     &                                & multi-rep  & 1        & \textbf{.644}$^{\dagger\ddagger}$ & \textbf{.458}$^{\dagger\ddagger}$ & \textbf{.596}$^{\dagger\ddagger}$ & .683$^{\dagger\ddagger}$ & .603$^{\dagger\ddagger}$ & .705$^{\dagger\ddagger}$ & .752 & .666 & .729 & .847 & .665 & \textbf{.830}$^\dagger$ \\
		&                                     & \multirow{2}{*}{LLaDA}         & single-rep & 1        & .620 & .446 & .579 & .640 & .595 & .674 & .733 & .681 & .743 & .840 & .691 & .823 \\
		&                                     &                                & multi-rep  & 1        & .622 & .452$^{\dagger\ddagger}$ & .584$^\ddagger$ & .647$^\dagger$ & \textbf{.613}$^{\dagger\ddagger}$ & .687$^\dagger$ & .744 & \textbf{.695}$^{\dagger\ddagger}$ & .746 & .846 & \textbf{.710}$^\ddagger$ & .819$^\ddagger$ \\
		\bottomrule
	\end{tabular}%
	}
\end{table}

\begin{table}[H]
	\ContinuedFloat
	\centering
	\scriptsize
	\setlength{\tabcolsep}{2pt}
	\caption{BEIR-7 D/S/H results, continued.}
	\resizebox{\textwidth}{!}{%
	\begin{tabular}{@{}clllccccccccccccc@{}}
		\toprule
		& \multirow{2}{*}{Method} & \multirow{2}{*}{Backbone} & \multirow{2}{*}{Variant} & \multirow{2}{*}{Fwd} & \multicolumn{3}{c}{FiQA} & \multicolumn{3}{c}{ArguAna} & \multicolumn{3}{c}{Quora} & \multicolumn{3}{c}{Avg} \\
		\cmidrule(lr){6-8} \cmidrule(lr){9-11} \cmidrule(lr){12-14} \cmidrule(lr){15-17}
		&        &          &         &     & D & S & H & D & S & H & D & S & H & D & S & H \\
		\midrule
		\multirow{11}{*}{\rotatebox[origin=c]{90}{\textbf{Zero-shot}}}
		& BM25                                & ---                            & sparse     & ---      & --- & \textbf{.236} & --- & --- & \textbf{.275} & --- & --- & \textbf{.789} & --- & --- & \textbf{.472} & --- \\
		\cmidrule(lr){2-17}
		& \multirow{2}{*}{DiffEmbed}          & Dream                          & single-rep & 1        & .143 & --- & --- & .308 & --- & --- & .685 & --- & --- & .300 & --- & --- \\
		&                                     & LLaDA                          & single-rep & 1        & .117 & --- & --- & .326 & --- & --- & .528 & --- & --- & .285 & --- & --- \\
		\cmidrule(lr){2-17}
		& \multirow{4}{*}{PromptReps}         & \multirow{2}{*}{Qwen2.5}       & single-rep & 1        & .200 & .193 & .271 & .189 & .179 & .209 & .693 & .667 & .784 & .321 & .401 & .441 \\
		&                                     &                                & multi-rep  & $\leq$20 & .165 & .188 & .239 & .231$^\dagger$ & .121 & .240$^\dagger$ & .600 & .634 & .707 & .342$^\dagger$ & .400$^\dagger$ & .446$^\dagger$ \\
		&                                     & \multirow{2}{*}{LLaMA3}        & single-rep & 1        & .272 & .205 & .318 & .229 & .173 & .248 & .729 & .682 & .804 & .412 & .421 & .503 \\
		&                                     &                                & multi-rep  & $\leq$20 & .212 & .207 & .270 & .315$^\dagger$ & .133 & .316$^\dagger$ & .538 & .686 & .772 & .389 & .424$^\dagger$ & .500 \\
		\cmidrule(lr){2-17}
		& \multirow{4}{*}{\textbf{DiffRetriever}} & \multirow{2}{*}{Dream}         & single-rep & 1        & .200 & .118 & .220 & .290 & .139 & .273 & .702 & .503 & .725 & .311 & .219 & .331 \\
		&                                     &                                & multi-rep  & 1        & .293$^{\dagger\ddagger}$ & .173$^\dagger$ & .278$^\dagger$ & .350$^{\dagger\ddagger}$ & .137 & .323$^{\dagger\ddagger}$ & .532 & .647$^\dagger$ & .696 & .427$^{\dagger\ddagger}$ & .376$^\dagger$ & .476$^{\dagger\ddagger}$ \\
		&                                     & \multirow{2}{*}{LLaDA}         & single-rep & 1        & \textbf{.311} & .218 & \textbf{.325} & .379 & .237 & \textbf{.370} & .811 & .706 & \textbf{.825} & .446 & .432 & .510 \\
		&                                     &                                & multi-rep  & 1        & .308$^\ddagger$ & .225$^\ddagger$ & .317 & \textbf{.386}$^\ddagger$ & .179 & .357$^\ddagger$ & \textbf{.819}$^{\dagger\ddagger}$ & .678 & .814$^\ddagger$ & \textbf{.497}$^{\dagger\ddagger}$ & .451$^{\dagger\ddagger}$ & \textbf{.539}$^{\dagger\ddagger}$ \\
		\midrule
		\multirow{11}{*}{\rotatebox[origin=c]{90}{\textbf{Fine-tuned}}}
		& RepLLaMA                            & LLaMA3                         & single-rep & 1        & .383 & --- & --- & .411 & --- & --- & .821 & --- & --- & .565 & --- & --- \\
		\cmidrule(lr){2-17}
		& \multirow{2}{*}{DiffEmbed}          & Dream                          & single-rep & 1        & .478 & --- & --- & .375 & --- & --- & .831 & --- & --- & .638 & --- & --- \\
		&                                     & LLaDA                          & single-rep & 1        & .455 & --- & --- & .387 & --- & --- & .842 & --- & --- & .595 & --- & --- \\
		\cmidrule(lr){2-17}
		& \multirow{4}{*}{PromptReps}         & \multirow{2}{*}{Qwen2.5}       & single-rep & 1        & .444 & .299 & .411 & .407 & .300 & .386 & .872 & .727 & .844 & .657 & .520 & .632 \\
		&                                     &                                & multi-rep  & $\leq$4  & .440 & .293 & .404 & .407 & .287 & .381 & .872 & .729 & .844 & .664$^\dagger$ & .530$^\dagger$ & .640$^\dagger$ \\
		&                                     & \multirow{2}{*}{LLaMA3}        & single-rep & 1        & .417 & .268 & .378 & .414 & \textbf{.311} & .391 & .860 & .715 & .834 & .651 & .512 & .623 \\
		&                                     &                                & multi-rep  & $\leq$4  & .437$^\dagger$ & .290$^\dagger$ & .395$^\dagger$ & \textbf{.416} & .290 & .390 & .865$^\dagger$ & .698 & .827 & .655$^\dagger$ & .526 & .633$^\dagger$ \\
		\cmidrule(lr){2-17}
		& \multirow{4}{*}{\textbf{DiffRetriever}} & \multirow{2}{*}{Dream}         & single-rep & 1        & .463 & .292 & .409 & .406 & .288 & .387 & .859 & .703 & .833 & .654 & .515 & .629 \\
		&                                     &                                & multi-rep  & 1        & \textbf{.479}$^{\dagger\ddagger}$ & \textbf{.316}$^{\dagger\ddagger}$ & \textbf{.431}$^{\dagger\ddagger}$ & .403 & .305$^\dagger$ & .382 & \textbf{.887}$^{\dagger\ddagger}$ & \textbf{.748}$^{\dagger\ddagger}$ & \textbf{.859}$^{\dagger\ddagger}$ & \textbf{.671}$^{\dagger\ddagger}$ & \textbf{.537}$^{\dagger\ddagger}$ & \textbf{.647}$^{\dagger\ddagger}$ \\
		&                                     & \multirow{2}{*}{LLaDA}         & single-rep & 1        & .453 & .302 & .406 & .414 & .294 & \textbf{.392} & .799 & .552 & .746 & .643 & .509 & .623 \\
		&                                     &                                & multi-rep  & 1        & .443$^\ddagger$ & .303$^\ddagger$ & .397$^\ddagger$ & .412 & .291 & .389 & .798 & .595$^\dagger$ & .760$^\dagger$ & .645 & .523$^\dagger$ & .626$^\dagger$ \\
		\bottomrule
	\end{tabular}%
	}
\end{table}

Table~\ref{tab:main_beir7} reports zero-shot results in
hybrid mode and fine-tuned results in dense mode, following the
scoring-mode shift observed in \S\ref{sec:results_finetuned}. Without
supervision, dense alone is less reliable than hybrid scoring; after
contrastive fine-tuning, dense scores improve enough that dense alone
usually exceeds hybrid. Table~\ref{tab:appendix_beir7_dsh} reports the
full BEIR-7 breakdown across dense, sparse, and hybrid scoring for every
system in both settings.

The in-domain mode flip transfers cleanly to BEIR-7. In the zero-shot
half, hybrid exceeds dense alone on every system on the BEIR-7 average,
by $0.020$ to $0.120$ NDCG@10. In the fine-tuned half, dense exceeds
hybrid on every system on the BEIR-7 average, by $0.019$ to $0.028$
NDCG@10. This supports the main-text choice to report hybrid for
zero-shot transfer and dense for fine-tuned transfer. The per-dataset
breakdown also shows where individual datasets favor sparse-heavy
behavior, such as COVID and Quora for some configurations.

\subsection{Additional Baseline Comparison}
\label{app:additional_comparison}

\begin{table}[H]
	\centering
	\scriptsize
	\setlength{\tabcolsep}{2pt}
\caption{Dense-only comparison with reported dense baselines and fine-tuned \method{} variants. Variant denotes retrieval representation count. In the MS~MARCO group, Dev uses MRR@10 and DL19/DL20 use NDCG@10; BEIR-7 uses NDCG@10. Contriever, TAS-B, and ColBERT-v2 use published scores.}
	\resizebox{\textwidth}{!}{%
	\begin{tabular}{@{}lll|ccc|cccccccc@{}}
		\toprule
		Method & Backbone & Variant & \multicolumn{3}{c|}{MS~MARCO} & \multicolumn{8}{c}{BEIR-7} \\
		\cmidrule{4-6}\cmidrule{7-14}
		& & & Dev & DL19 & DL20 & NQ & HQA & SciFact & COVID & FiQA & ArguAna & Quora & Avg \\
		\midrule
		Contriever & BERT & single-rep & .341 & .678 & .661 & .498 & .638 & .677 & .596 & .329 & .446 & .865 & .578 \\
		TAS-B & BERT & single-rep & .340 & .712 & .693 & .465 & .584 & .644 & .505 & .296 & .436 & .835 & .538 \\
		ColBERT-v2 & BERT & multi-rep & .397 & .744 & .740 & .562 & .667 & .693 & .738 & .356 & .463 & .852 & .619 \\
		\midrule
		Qwen3-Embed-0.6B & Qwen3-0.6B & single-rep & .315 & .662 & .669 & .535 & .651 & .702 & .883 & .458 & .481 & .869 & .654 \\
		Qwen3-Embed-4B & Qwen3-4B & single-rep & .357 & .750 & .732 & .632 & .717 & .768 & .884 & .585 & .504 & .876 & .710 \\
		Qwen3-Embed-8B & Qwen3-8B & single-rep & .369 & \textbf{.757} & .740 & .652 & .737 & .788 & \textbf{.929} & .611 & \textbf{.524} & .877 & .731 \\
		NV-Embed-v2 & Mistral-7B & single-rep & .390 & .751 & \textbf{.764} & \textbf{.729} & \textbf{.823} & \textbf{.807} & .886 & \textbf{.649} & .478 & .886 & \textbf{.751} \\
		GritLM-7B & Mistral-7B & single-rep & .343 & .703 & .708 & .599 & .766 & .781 & .743 & .588 & .435 & \textbf{.892} & .686 \\
		PPLX-Embed-4B & PPLX & single-rep & .364 & .717 & .735 & .632 & .737 & .746 & .857 & .574 & .487 & .887 & .703 \\
		\midrule
		\multirow{4}{*}{\textbf{DiffRetriever}} & \multirow{2}{*}{Dream} & single-rep & .424 & .741 & .721 & .619 & .650 & .739 & .841 & .463 & .406 & .859 & .654 \\
		&  & multi-rep & \textbf{.433} & .751 & .729 & .644 & .683 & .752 & .847 & .479 & .403 & .887 & .671 \\
		\cmidrule{2-14}
		& \multirow{2}{*}{LLaDA} & single-rep & .424 & .715 & .715 & .620 & .640 & .733 & .840 & .453 & .414 & .799 & .643 \\
		&  & multi-rep & .427 & .718 & .721 & .622 & .647 & .744 & .846 & .443 & .412 & .798 & .645 \\
		\bottomrule
	\end{tabular}%
	}
	\label{tab:dense-ft-baselines}
\end{table}

Table~\ref{tab:dense-ft-baselines} reports a broader comparison
against dense retrieval baselines beyond the matched-recipe baselines
used in the main text. We group the baselines into BERT-era dense
retrievers (Contriever, TAS-B, ColBERT-v2) and recent LLM-based
dedicated embedding models (the Qwen3-Embedding line, NV-Embed-v2,
GritLM-7B, and PPLX-Embed-4B).

The LLM-based embedding models are not controlled comparisons for the
\method{} retrieval mechanism. They are dedicated embedding models
trained on substantially broader retrieval data, including multi-dataset
mixtures, MTEB-style instruction-tuning data, synthetic queries,
curated hard negatives, and multi-stage training pipelines. \method{},
in contrast, is fine-tuned only on MS~MARCO Tevatron triples
(\S\ref{sec:exp_finetune}). We therefore report these models as
landscape context rather than matched baselines.

Three patterns emerge. First, in the in-domain MS~MARCO dev setting,
multi-representation \method{} with Dream achieves the highest score
($.433$), exceeding every dedicated embedding baseline in the table,
including NV-Embed-v2 ($.390$), Qwen3-Embedding-8B ($.369$), and
Qwen3-Embedding-4B ($.357$). Second, \method{} substantially
outperforms all BERT-era dense retrievers on the BEIR-7 average,
including ColBERT-v2 ($.619$ vs.\ Dream multi $.671$). Third, on
BEIR-7, dedicated embedding models trained on broader retrieval
mixtures, such as Qwen3-Embedding and NV-Embed-v2, achieve higher
averages than \method{}, consistent with their broader out-of-domain
training coverage. Fine-tuning \method{} on broader retrieval mixtures
is a natural next step.


\section{Supplementary Analysis}
\label{app:analysis}

This section provides additional analyses of representation-budget
selection and adaptive-budget headroom.

\subsection{Per-Dataset (\texorpdfstring{$K_q, K_p$}{Kq, Kp})
	Landscape}
\label{app:per_dataset_heatmaps}

\begin{figure*}[t]
	\centering
	\includegraphics[width=\textwidth]{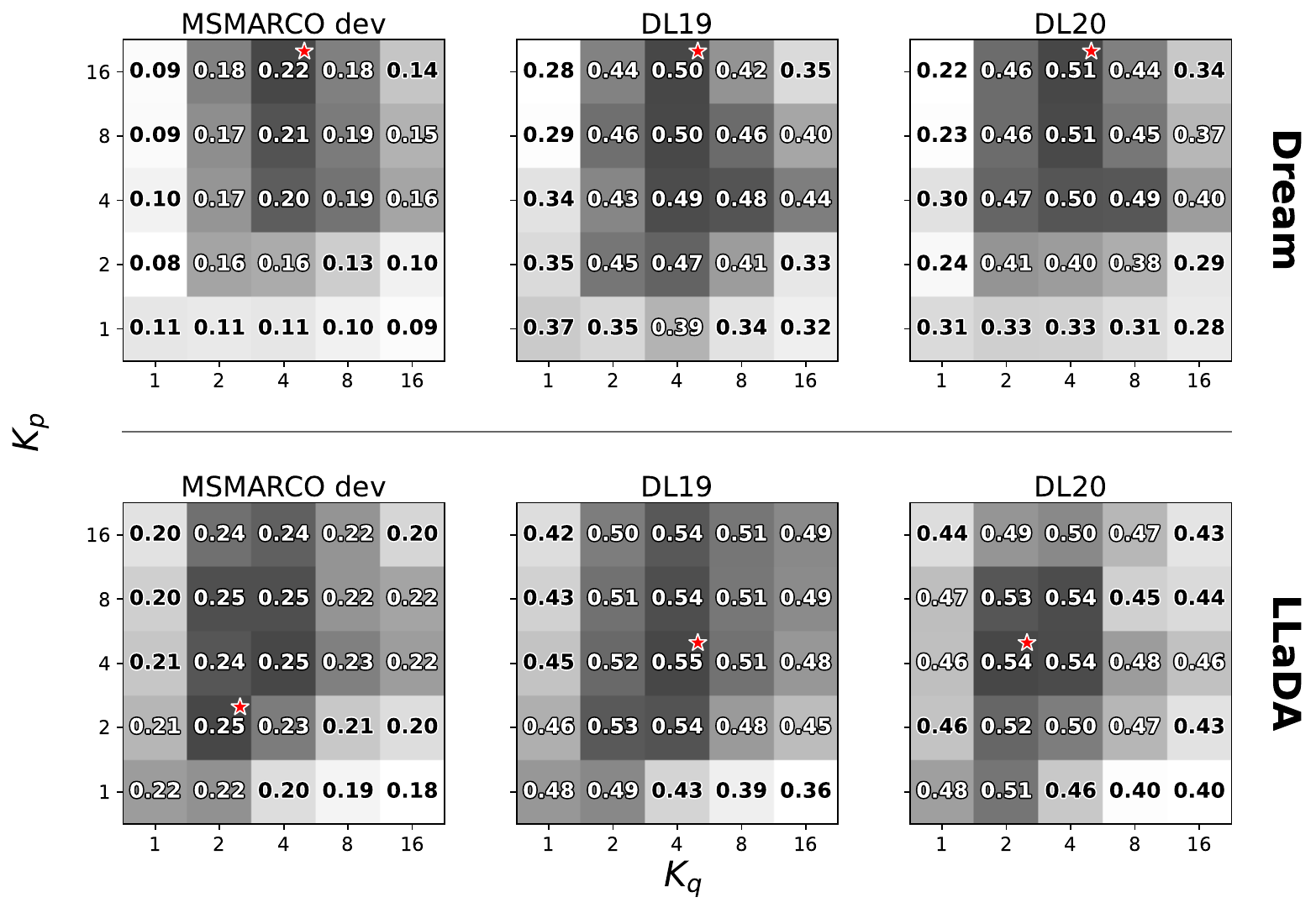}
	\caption{In-domain per-dataset zero-shot hybrid retrieval
		landscape on MS~MARCO dev (MRR@10), TREC DL19, and TREC
		DL20 (NDCG@10), across $(K_q,K_p) \in \{1,2,4,8,16\}^2$,
		for Dream (top row) and LLaDA (bottom row). Stars mark
		the per-panel best-performing cell. The train-selected
		$(K_q^{*}, K_p^{*})$ used in the main text is $(4,16)$
		for Dream and $(4,4)$ for LLaDA. MS~MARCO dev
		reproduces the corresponding panel from
		Figure~\ref{fig:kqkp_heatmap}.}
	\label{fig:per_dataset_heatmaps_indomain}
\end{figure*}

\begin{figure*}[t]
	\centering
	\includegraphics[width=\textwidth]{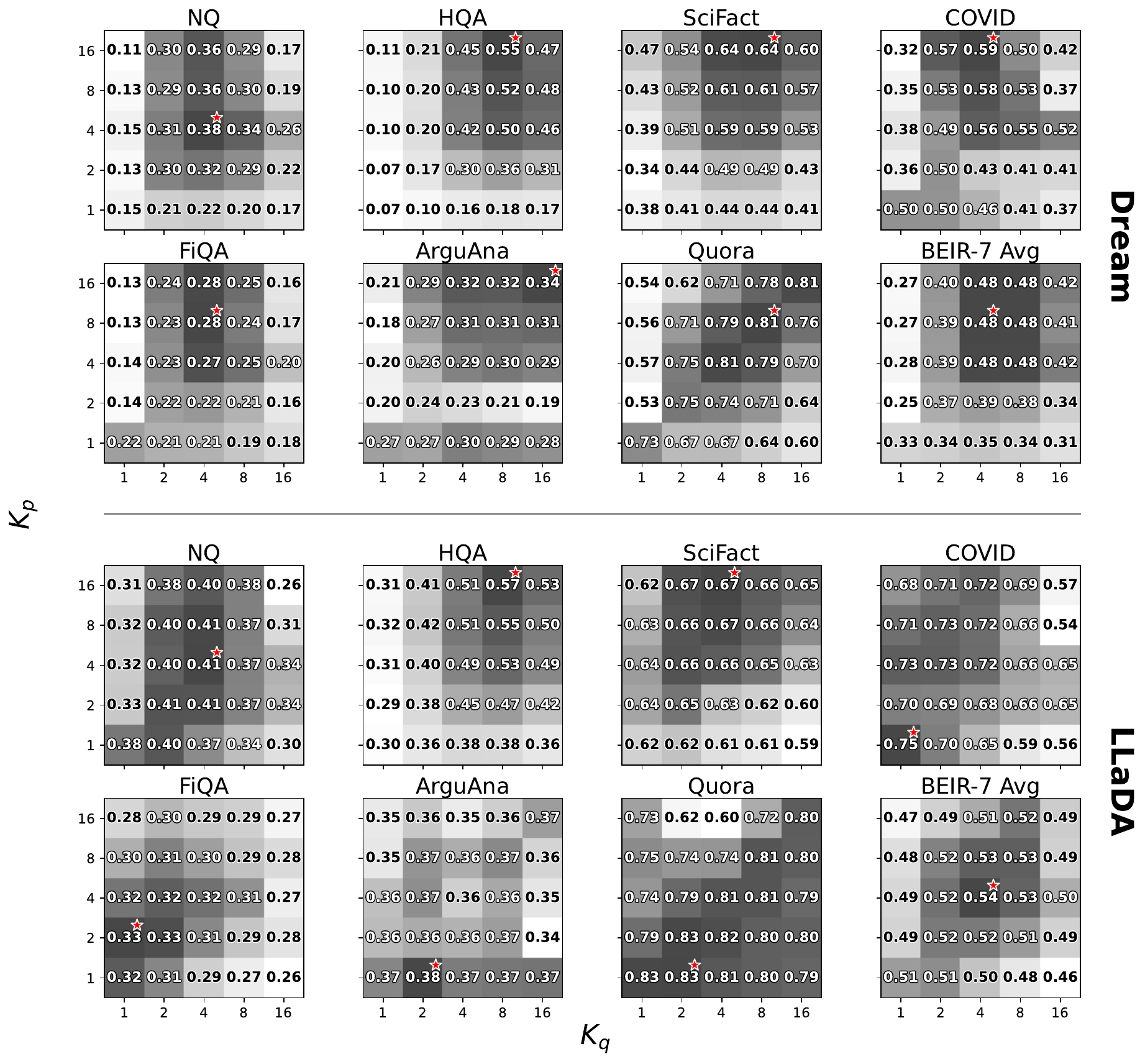}
	\caption{Out-of-domain per-dataset zero-shot hybrid retrieval
		landscape on the seven BEIR-7 datasets and their average
		(NDCG@10), across $(K_q,K_p) \in \{1,2,4,8,16\}^2$, for
		Dream (top two rows) and LLaDA (bottom two rows). Stars
		mark the per-panel best-performing cell. The
		train-selected $(K_q^{*}, K_p^{*})$ used in the main
		text is $(4,16)$ for Dream and $(4,4)$ for LLaDA. The
		Avg panel reproduces the BEIR-7 average from the
		Figure~\ref{fig:kqkp_heatmap} as a
		within-figure reference.}
	\label{fig:per_dataset_heatmaps_beir7}
\end{figure*}

Figure~\ref{fig:kqkp_heatmap} reports the
$(K_q,K_p)$ landscape on MS~MARCO dev and the BEIR-7 average.
Here, we expand this view to individual datasets: in-domain DL19
and DL20 (Figure~\ref{fig:per_dataset_heatmaps_indomain}), and
the seven BEIR-7 datasets
(Figure~\ref{fig:per_dataset_heatmaps_beir7}). Both figures also
include the corresponding aggregate panel from the main text as a
within-figure reference.

The in-domain panels show a consistent pattern. For Dream, the
train-selected $(4,16)$ lands at the per-panel best on every
in-domain dataset. For LLaDA, the train-selected $(4,4)$ lands at
the per-panel best on DL19, and on the other two in-domain panels
it lies within a flat plateau where multiple cells tie. The
per-backbone asymmetry from \S\ref{sec:analysis_landscape}---Dream
passage-heavy, LLaDA more isotropic around the symmetric
center---holds in every in-domain panel.

The out-of-domain picture is more varied. The train-selected cell
lands at the best cell on the BEIR-7 average panel for both
backbones, but per-dataset peaks scatter across the grid. HotpotQA
peaks at $(8,16)$ for both backbones, consistent with multi-hop QA
benefiting from larger query and passage budgets. LLaDA on
TREC-COVID peaks at $(1,1)$, while LLaDA on ArguAna and Quora peaks
at the $K_p=1$ row, suggesting that additional passage
representations help less for these datasets. The train-selected
cell is rarely more than $.02$ below the per-panel best on most
datasets, but the preferred budgets differ substantially.

This variance is consistent with the per-query oracle analysis in
\S\ref{sec:analysis_oracle}: a fixed $(K_q^{*},K_p^{*})$
works well in aggregate but leaves effectiveness on the table when
queries or datasets prefer different budgets.

\end{document}